%% file: main.tex
\documentclass[letterpaper]{article} 
\usepackage{aaai24}  
\usepackage{times}  
\usepackage{helvet}  
\usepackage{courier}  
\usepackage[hyphens]{url}  
\usepackage{graphicx} 
\usepackage{natbib}  
\usepackage{caption} 
\usepackage{algorithm}
\usepackage{algpseudocode}

\usepackage{newfloat}
\usepackage{listings}
\DeclareCaptionStyle{ruled}{labelfont=normalfont,labelsep=colon,strut=off} 
\floatstyle{ruled}
\newfloat{listing}{tb}{lst}{}
\floatname{listing}{Listing}
\title{

Physics-Informed Representation and Learning: Control and Risk Quantification
}
\author{
    Zhuoyuan Wang\textsuperscript{\rm 1},
    Reece Keller\textsuperscript{\rm 1},
    Xiyu Deng\textsuperscript{\rm 1}, Kenta Hoshino\textsuperscript{\rm 2}, Takashi Tanaka\textsuperscript{\rm 3}, Yorie Nakahira\textsuperscript{\rm 1}
}
\affiliations{
    \textsuperscript{\rm 1}Carnegie Mellon University \\
    \textsuperscript{\rm 2}Kyoto University \\
    \textsuperscript{\rm 3}University of Texas at Austin
    \\

    \{zhuoyuaw, rdkeller, xiyud\}@andrew.cmu.edu, hoshino@i.kyoto-u.ac.jp, ttanaka@utexas.edu, yorie@cmu.edu
}

\usepackage{xcolor}
\usepackage{amsmath}
\usepackage{amsthm}
\usepackage{amsfonts}
\usepackage{graphicx}
\usepackage{caption}
\usepackage{subcaption}
\usepackage{bm}

\newcommand{\resolved}[1]{}

\newcommand{\ie}{\textit{i.e., }}
\newcommand{\eg}{\textit{e.g., }}
\newcommand{\pr}{\mathbb{P}}
\newcommand{\safe}{\mathcal{C}}

\newtheorem{theorem}{Theorem}

\newtheorem{definition}[theorem]{Definition}
\newtheorem{remark}[theorem]{Remark}
\newtheorem{assumption}[theorem]{Assumption}

\usepackage{tikz}
\usepackage{xcolor}
\usepackage{verbatim}
\usepackage{multirow}
\usepackage{color}
\usepackage{colortbl}
\usepackage{pgfplots}
\DeclareUnicodeCharacter{2212}{−}
\usepgfplotslibrary{groupplots,dateplot}
\usetikzlibrary{patterns,shapes.arrows}
\pgfplotsset{compat=newest}
\usetikzlibrary{shapes.arrows}
\usetikzlibrary{arrows.meta} 
\usetikzlibrary{positioning, fit, shapes.geometric}
\usetikzlibrary{backgrounds}
\definecolor{newblue}{RGB}{189,215,239} 
\definecolor{newgray}{RGB}{214,214,214} 
\definecolor{neworange}{RGB}{241,205,177} 
\definecolor{newyellow}{RGB}{251,231,163} 
\definecolor{newgreen}{RGB}{202,223,184} 
\definecolor{cmured}{RGB}{196,18,48} 
\definecolor{cmuskibored}{RGB}{147,17,32} 
\definecolor{cmuyellow}{RGB}{253,181,21} 
\definecolor{cmublue}{RGB}{4,54,115} 
\definecolor{cmutan}{RGB}{188,180,158} 
\definecolor{cmuteal}{RGB}{31,76,76} 
\definecolor{cmugreen}{RGB}{113,159,148} 
\definecolor{cmuskyblue}{RGB}{0,123,192} 
\definecolor{cmuirongray}{RGB}{109,110,113} 
\definecolor{cmugray}{RGB}{244,244,244} 
\usepackage{pgfplots}
\usepackage{makecell}
\usepgfplotslibrary{fillbetween}

\begin{document}

\maketitle

\begin{abstract}
Optimal and safety-critical control are fundamental problems for stochastic systems, and are widely considered in real-world scenarios such as robotic manipulation and autonomous driving. In this paper, we consider the problem of efficiently finding optimal and safe control for high-dimensional systems. Specifically, we propose to use dimensionality reduction techniques from a comparison theorem for stochastic differential equations together with a generalizable physics-informed neural network to estimate the optimal value function and the safety probability of the system. The proposed framework results in substantial sample efficiency improvement compared to existing methods. We further develop an autoencoder-like neural network to automatically identify the low-dimensional features of the system to enhance the ease of design for system integration. We also provide experiments and quantitative analysis to validate the efficacy of the proposed method. 
Source code is available at {https://github.com/jacobwang925/path-integral-PINN}.

\end{abstract}

\vspace{-0.4em}

\section{Introduction}

Optimal control and safety-critical control are the two central concerns for autonomous systems.
These concerns are particularly pronounced in real-world applications, \eg manufacturing robots and autonomous cars.
The operational environment often introduces stochastic noise, which compounds the difficulties of achieving optimal performance and ensuring safety. Traditional deterministic methods prove inadequate for stochastic dynamics~\cite{katsoulakis2020data}.
Furthermore, many real-world systems are characterized by high-dimensional state spaces (\eg multi-agent systems), leading to substantial computational burdens when devising optimal and safe control strategies. 


Previous work on stochastic optimal control deals with diverse uncertainty and randomness, but these methods are not efficient for high-dimensional systems because forward rollouts and backward dynamic programming requires computation that scales exponentially with the dimension of the state~\cite{gorodetsky2018high, frankowska2020necessary}\resolved{to elaborate}.
Previous studies on stochastic safe control aim to control the level of risk in the system and ensure the probability of safety does not decay over time~\cite{samuelson2018safety,gomez2016real}.
These methods rely on accurate estimations of the probability of risk in the system, and standard methods for such risk estimation are computationally heavy, especially for high-dimensional systems.
For both problems, existing methods require computation that scales exponentially or linearly with the time horizon of interest. To the best of our knowledge, there is no study that provides estimation of the value function or risk probability in long time horizons without introducing additional computation, which can be very beneficial for systems with long-term performance requirements.

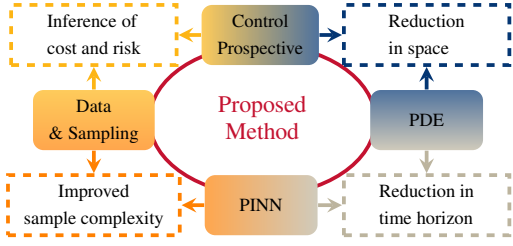
\begin{figure}
    \centering
    \input{Figures/overall_diagram}
    \caption{The overall diagram of the proposed method for stochastic optimal value function and safety probability estimation. \resolved{remove task x.x and replace with content in this paper}
    }
    \vspace{-1em}
    \label{fig:overall_diagram}
\end{figure}

In order to address these challenges in computation, this study proposes a unified framework to efficiently estimate the value function and safety probability of high-dimensional stochastic systems. The method leverages a comparison theorem~\cite{10.1215/kjm/1250523321} to find low-dimensional representations of the value function and safety probability, and uses such low-dimensional features to construct low-dimensional partial differential equations (PDEs) for the value function and safety probability calculation in order to reduce the dimension of the problem for efficient computation.
We further propose a physics-informed neural network (PINN) to solve these PDEs for better sample complexity and generalization abilities.
We also propose an autoencoder-like neural network to automatically identify low-dimensional features of the system.
Fig.~\ref{fig:overall_diagram} shows the overall diagram of the proposed method. 
The advantages of the proposed method are
\begin{enumerate}
    \item A unified framework for accurate estimation of value functions and safety probabilities of stochastic systems (Fig.~\ref{fig:PINN diagram}). 
    \item Efficient estimation with much lower sample complexity for high-dimensional systems (Fig.~\ref{fig:sample_complexity}).
    \item Generalization to unseen regions in the state space and longer time horizons 
    (Fig.~\ref{fig:value_func_visual_1000d}).
    \item Intuitive plug and use with automatic feature identification (Fig.~\ref{fig:AE diagram}).
\end{enumerate}

\vspace{-0.4em}



\vspace{-0.5em}

\section{Related Work}
\label{sec:related_work}
\subsection{Path Integral Optimal Control}
Path integral control generally refers to numerical methods to solve a stochastic optimal control problem by repeatedly 
performing forward Monte Carlo rollouts of open-loop dynamics. The original derivation of the path integral control algorithm~\cite{kappen2005path} relied on the Feynman-Kac lemma, whereas an alternative derivation without the Feynman-Kac lemma became available later~\cite{theodorou2012relative} based on the variational approach. In both derivations, the path integral method is restricted to the class of stochastic optimal control problems whose Hamilton–Jacobi–Bellman (HJB) PDEs are linearizable, but generalizations have been considered in~\cite{satoh2016iterative,williams2017information}. Various implementations such as path integral for policy improvements~\cite{theodorou2010generalized} and receding horizon implementations~\cite{williams2017model} have been widely used. Path integral control has also been applied to constrained systems and systems with non-differentiable dynamics~\cite{satoh2020nonlinear,carius2022constrained}. Path integral control for risk minimization in robot navigation has been considered in~\cite{patil2022chance}, which requires solving PDEs whose dimension scales with the size of the system. Here, we show that the value function and safety probability can be bounded exactly by the solution of low-dimensional PDEs regardless of the system dimension. 
\subsection{Risk Quantification for Safe Control}
\resolved{similar topic but alternative objectives, or just cite related safe control papers}
\resolved{add more recent papers on risk quantification}

Risk quantification is the key enabler for many long-term stochastic safe control methods~\cite{wang2021myopically,jing2022probabilistic}.
Existing methods often use rare event simulation through Monte Carlo (MC) and importance sampling to estimate the long-term risk in stochastic systems~\cite{ botev2013markov,hanna2021importance,stadie2018some,madhushani2020hamiltonian}. The subset simulation calculates the risk probability conditioned on intermediate failure events for improved sample efficiency~\cite{huang2016assessing, zhao2022subset, rashki2021sesc}. Probabilistic reachability estimates the risk of controllers in stochastic systems by propagating the estimated risk backwards over time~\cite{hewing2018stochastic,bansal2020hamilton,huh2020safe}. MC techniques typically require samples to cover states and evaluate the risk over the time horizon.
PDE techniques are also used to understand probabilistic values in stochastic systems~\cite{chern2021safe,feng2014comparative}, but numerical PDE techniques such as finite difference, finite element, and finite volume methods are less scalable than MC methods. 
Probability bounds and martingale inequalities have been used to approximate risk probabilities for certain classes of systems~\cite{clark2019control,yaghoubi2020risk,santoyo2021barrier,cheng2020safe, meng2022sufficient, nishimura2023control}. 
The large deviation is another standard approach that can be adapted to the safe control area, which allows evaluating the probability of the state of a stochastic differential equation that exists from a given region~\cite{bressloff2014path,bertini2015large}. 
Nonetheless, most existing methods suffer from the curse of dimensionality. In this work, we solve the high-dimensional safety probability estimation problem by finding effective low-dimensional representations, and we are able to reduce the sample complexity by orders compared to MC methods.


\subsection{Physics-informed Neural Networks}
Physics-informed neural networks (PINNs) are neural networks that are trained to solve supervised learning tasks while respecting any given laws of physics described by general nonlinear PDEs~\citep{raissi2019physics}. PINNs take both data and the physics model of the system into account, and are able to solve the forward problem of getting PDE solutions, and the inverse problem of discovering underlying governing PDEs from data. PINNs have been widely used in power systems~\cite{misyris2020physics}, fluid mechanics,~\cite{cai2022physics} medical care~\cite{sahli2020physics}, and risk quantification~\cite{han2018solving, pereira2021safe, wang2023generalizable}. We leverage the generalization ability of PINNs to efficiently estimate value functions and safety probabilities in unseen regions of the state space to further enhance the sample complexity of the proposed method.

\vspace{-0.1em}

\section{Problem Formulation}
\label{sec:problem_formulation}

\subsection{System Dynamics}
Consider the following class of nonlinear stochastic dynamical systems defined on a probability space $(\Omega,F,P)$:
\begin{equation}
\label{eq:continuous_dynamics}
\begin{aligned}
d x_t & =f\left(x_t\right) d t+\sigma\left(x_t\right) \left(u_t dt + dw_t\right)
\end{aligned}
\end{equation}
Here, $x_t \in \mathcal{X} \subset \mathbb{R}^n$, $0 \leq t \leq T$ is the state of the system and $u_t \in \mathcal{U} \subset \mathbb{R}^m$, $0 \leq t \leq T$ is the control input, $w_t$ is the $m$-dimensional standard Brownian motion in the probability measure $P$ and $\sigma$ is the diffusion coefficient.
The role of the controller is to apply the control input $u_t$ based on a state feedback policy \ie $u_t$ is measurable with respect to the filtration $F^{x_t}$ generated by $\{x_\tau\}_{0\leq \tau\leq t}$. Suppose that $Q$ is an alternative probability measure in which $u_t \equiv 0$. Correspondingly, we have
\begin{equation}
    d\tilde{w}_t=u_tdt+dw_t, \tilde{w}_0=0
\end{equation} 
is the standard Brownian motion.

\subsection{Stochastic Optimal Control}
Consider the running cost defined as
\begin{equation}
\label{eq:quadratic_cost_function}
    w(x_t, u_t)=c(x_t)+\frac{1}{2}\|u_t\|^2,
\end{equation}
where $c: \mathbb{R}^n \rightarrow \mathbb{R}$. 
The stochastic optimal control problem aims to find the optimal value function
\begin{equation}
\label{eq:value_function}
    V(x, t):=\min_{u}\mathbb{E}^{P}\left[\int_{t}^{T} w(x_\tau, u_\tau) d \tau+c\left(x_{T}\right) \mid x_{t}=x\right],
\end{equation}
which explicitly yields the optimal control  as
\begin{equation}
\label{eq:optimal_control_from_V}
    u_t^{\star}=-\sigma\left(x_t\right)^\top \nabla_x V\left(x_t, t\right).
\end{equation}
A notable feature of the stochastic optimal control problems is the applicability of Monte Carlo-based numerical solution strategy, which we call the path-integral method~\cite{thijssen2015path}. For each time $t \in[0, T)$ and the state $x \in \mathbb{R}^{n}$, the path-integral method allows the control agent to compute the optimal input $u_t^{\star}$ by evaluating the path integrals along randomly generated state trajectories $x_{\tau}, t \leq \tau \leq T$ starting from $x_{t}=x$.
From existing results on KL control and free energy~\cite{fleming2006controlled, boue1998variational, theodorou2012relative}, the value function~\eqref{eq:value_function} can be solved explicitly as
\begin{equation}
\label{eq:value_function_free_energy}
\begin{aligned}
    & V(x, t) = \\
    & -\log \mathbb{E}^{Q}\left[\exp \left(- \int_{t}^{T} c\left(x_{\tau}\right) d \tau- c\left(x_{T}\right)\right) \mid x_{t}=x\right].
\end{aligned}
\end{equation}
Since the right hand side of~\eqref{eq:value_function_free_energy} contains the expectation with respect to $Q$, one can consider approximating it by Monte Carlo simulation as
\begin{equation}
\label{eq:path_integral_MC_estimation}
    V(x, t) \approx 
     - \log \left[\frac{1}{N} \sum_{i=1}^{N} \exp \left(- \int_{t}^{T} c\left(x_{\tau}^{i}\right) d \tau- c\left(x_{T}^{i}\right)\right)\right],
\end{equation}
where $\left\{x_{\tau}^{i}, t \leq \tau \leq T\right\}_{i=1}^{N}$ are randomly drawn sample paths from distribution $Q$. 
Since $\tilde{w}_t$ is the standard Brownian motion under $Q$, such sample paths can be obtained by simply simulating the uncontrolled system $dx_t = f(x_t)dt + \sigma(x_t) d\tilde{w}_t$.
Note that $Q$ is the uncontrolled process and is easy to sample from, while the value function given by~\eqref{eq:path_integral_MC_estimation} is associated with the optimal control.
This is the path integral control method, and is widely adopted for low-dimensional systems. However, when the system dimension is high, sampling~\eqref{eq:path_integral_MC_estimation} is nontrivial. We aim to address this issue with the proposed framework.


\subsection{Safety-critical Control}
We consider system~\eqref{eq:continuous_dynamics} with a nominal control policy $u = \mathcal{N}(x)$. We define safety of the system as the state staying within a safe set $\mathcal{C}$, which is the super-level set of a barrier function $\phi(x): \mathbb{R}^n \rightarrow \mathbb{R}$, \ie 
\begin{equation}
\label{eq:safe set definition}
    \mathcal{C} = \{x : \phi(x) \geq 0\}.
\end{equation}
This definition of safety can characterize a large variety of practical safety requirements~\cite{prajna2007framework, ames2019control}. Since in stochastic systems safety can only be guaranteed in the sense of probability, we consider long-term safety probability $F$ of the system defined as below.
\begin{definition}[Safety probability]
\label{def:safety probability}
Starting from initial state $x_0 = x \in \safe$, the safety probability $F$ of system~\eqref{eq:continuous_dynamics} for outlook time horizon $t$ is defined as the probability of state $x_\tau$ staying in the safe set $\safe$ over the time interval $[0,t]$, i.e., 
\begin{equation}
    F(x,t) = \pr(x_\tau \in \safe, \forall \tau\in [0,t] \mid x_0 =x).
\end{equation}
\end{definition}
The goal is to find the safety probability $F$ over the state space for a long-term horizon $T$. Once the safety probability is acquired, existing safe control methods can be used to guarantee safety of the system~\cite{wang2021myopically,jing2022probabilistic}.
One standard approach to acquire such safety probability is to run Monte Carlo simulation with the nominal controller $\mathcal{N}$ multiple times and calculate the empirical probability of the system being safe, \ie
\begin{equation}
    \bar{F}(x,T) = \frac{N_\text{safe}}{N} \approx \pr(x_\tau \in \mathcal{C}, \forall \tau \in [0,T] \mid x_0 = x),
\end{equation}
where $N_\text{safe}$ is the number of safe trajectories over $N$ trajectories. However, such estimation has a sample complexity that scales linearly with horizon $T$ and exponentially with system dimension $n$~\cite{rubino2009rare, wang2021myopically}, and is thus inefficient for high-dimensional systems with long-term safety requirements. We aim to overcome these issues and efficiently estimate long-term safety probabilities of high-dimensional systems with the proposed framework.



\section{Proposed Method}
\label{sec:proposed_method}
\begin{figure}[t]
    \centering
    \input{Figures/flowchart}
    \caption{The procedure diagram of the proposed method.}
    \label{fig:flowchart}
\end{figure}
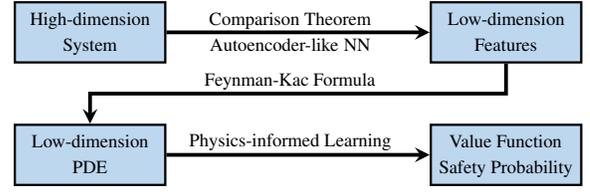

In the section, we introduce the proposed framework to efficiently estimate value functions and safety probabilities of high-dimensional systems. The method consists of three procedures. We first apply comparison theorem to find low-dimensional features of the system and the associated stochastic processes that characterize their evolution. 
Then we transform the stochastic process into the solution of certain PDEs. 
Last, we formulate a physics-informed learning problem to efficiently solve the PDE with special boundary and initial conditions. Additionally, we introduce an autoencoder-like neural network for automatic feature identification. Fig.~\ref{fig:flowchart} shows the overall procedures of the proposed framework.

\subsection{Comparison Theorem for Feature Identification}
We assume the low-dimensional feature can be represented by a smooth function $p: \mathbb{R}^n \rightarrow \mathbb{R}$. Here, we use comparison theorem~\cite[Theorem 3.1]{ikeda1977comparison} to find the stochastic process $\xi$ that describes the exact evolution of $p(x)$.
We introduce the operator $\mathcal{A}$ as
\begin{equation}
\label{eq:def_L}
\begin{aligned}
\mathcal{A}^U(\cdot)(x) & =\frac{\partial(\cdot)}{\partial x}(x) f(x)+ \frac{\partial (\cdot)}{\partial x}(x) \sigma(x) U + \\
& \qquad \frac{1}{2} \operatorname{Tr}\left(\frac{\partial^2(\cdot)}{\partial x^2}(x) \sigma(x) \sigma(x)^\top\right),
\end{aligned}
\end{equation}
where $U: \mathbb{R}^n \rightarrow \mathbb{R}^m$ is the control policy that maps the state to the control input, and we define
\begin{equation}
    a(x)=\sum_{i, j, k} \sigma_k^i(x) \sigma_k^j(x) \frac{\partial p}{\partial x_i}(x) \frac{\partial p}{\partial x_j}(x), \; b(x)=\frac{\mathcal{A}^U p(x)}{a(x)},
\end{equation}
where $\sigma_k^i(x)$ is the $(i,k)$-th element of the diffusion coefficient $\sigma(x)$.
For value function estimation, we consider $\mathcal{A}^0$ for the uncontrolled process where $U \equiv 0$, and for safety probability estimation, we consider $\mathcal{A}^{\mathcal{N}}$ where $U = \mathcal{N}$ is the nominal control policy.
Additionally, let $\xi \in I \subset \mathbb{R}$ be a scalar variable, where $I$ is the range of the feature $p(x)$. Using similar notation employed in~\cite{ikeda1977comparison}, consider
\begin{equation}
\label{eq:a_b_def}
\begin{aligned}
    a^{+}(\xi) = \sup _{x: p(x)=\xi} a(x), & \quad a^{-}(\xi) = \inf _{x: p(x)=\xi} a(x) \\
    b^{+}(\xi)= \sup _{x: p(x)=\xi} b(x), & \quad b^{-}(\xi)=\inf _{x: p(x)=\xi} b(x)
\end{aligned}
\end{equation}
\begin{assumption}
\label{asm:match_upper_lower_bound}
We assume that the feature function $p(x)$ satisfies that $a^{+}(\xi) = a^{-}(\xi) = \alpha(\xi)$ and $b^{+}(\xi) = b^{-}(\xi) = \beta(\xi)$, $\forall \xi \in I$. 
\end{assumption}
\begin{assumption}
\label{asm:continuity-condition-on-a-and-b}
The functions $\alpha(\xi)$ and $\beta(\xi)$ are globally Lipschitz continuous in $\xi \in I \subset \mathbb{R}$.
Moreover, $a(x) > 0$, $\forall x \in \mathbb{R}^{n}$.
\end{assumption}

\begin{theorem}
\label{thm:comparison_lemma}
Given Assumptions~\ref{asm:match_upper_lower_bound} and~\ref{asm:continuity-condition-on-a-and-b} hold, $p(x_t)$ with $x_t$ being sampled from system~\eqref{eq:continuous_dynamics} is characterized by the following stochastic process
\begin{equation}
\label{eq:one_dim_xi_process}
    d \xi_t = \alpha\left(\xi_t\right) \beta\left(\xi_t\right) d t + \sqrt{\alpha\left(\xi_t\right)} d \tilde{B}_t,
\end{equation}
with $\xi_0 = p(x_0)$, and $\tilde{B}_t$ being a one-dimensional standard Wiener process.
\end{theorem}

\vspace{0.2em}

\noindent \textit{Proof.} \; See Appendix.

\vspace{0.3em}


Assumption~\ref{asm:match_upper_lower_bound} gives the conditions for the upper and lower bounds to match with the actual value $\alpha$ and $\beta$, thus a single stochastic process $\xi$ can be derived.
Until here, we have found the one-dimensional process~\eqref{eq:one_dim_xi_process} that characterizes the evolution of the feature of the high-dimensional system without any information loss.

\subsection{PDE for Value Function with Feynman-Kac}
\label{sec:Feynman_Kac_PDE}
In this section, we describe how to estimate the value function with the solution of a low-dimensional PDE.
We consider $p(x) = c(x)$, \ie the feature of the high-dimensional system is the value of the running cost on state.
We set
\begin{equation}
\label{eq:V_exp_transformation}
    V(x, t)=-\log \varphi(x, t).
\end{equation}
Let $\xi = p(x)$, then from~\eqref{eq:value_function_free_energy} we have
\begin{equation}
\label{eq:xi_value_function}
\begin{aligned}
    \varphi(x, t) 
    & = \mathbb{E} \left[\exp \left(-\int_t^T \xi_\tau d \tau - \xi_T\right)  \mid \xi_t=\xi\right] 
\end{aligned}
\end{equation}
With that, we apply Feynman-Kac formula~\cite{del2004feynman} on~\eqref{eq:xi_value_function} and~\eqref{eq:one_dim_xi_process} and get $\varphi(x, t)$ is the solution of the following two-dimensional PDE
\begin{equation}
\label{eq:one_dim_PDE}
\begin{aligned}
    W_{\varphi}(\xi,t) := & \frac{\partial \varphi}{\partial t}(\xi, t) + \alpha(\xi, t)\beta(\xi, t)\frac{\partial \varphi}{\partial \xi}(\xi, t) \\
    & +\frac{1}{2} \alpha(\xi, t) \frac{\partial^2 \varphi}{\partial \xi^2}(\xi, t) - \xi \varphi(\xi, t)=0,
\end{aligned}
\end{equation}
with initial (terminal) condition
\begin{equation}
\label{eq:one_dim_PDE_ic}
    \varphi(\xi, T) = \exp(-\frac{1}{2}\xi).
\end{equation}

\subsection{PDE for Safety Probability}
In this section, we describe how to estimate the safety probability with the solution of a low-dimensional PDE. We consider $p(x) = \phi(x)$, \ie the feature of the high-dimensional system is the value of the barrier function for the state. Then we have
\begin{equation}
\label{eq:safety_prob_hitting_time}
\begin{aligned}
    & \quad F(x,t) = \pr\left(x_\tau \in \safe, \forall \tau\in [0,t] \mid x_0 =x \right) \\
    & = \pr\left(\min_{0\leq \tau \leq t} \phi(x_\tau) \geq 0\right) = \pr\left(\min_{0\leq \tau \leq t} p(x_\tau) \geq 0\right) \\
    & = \pr\left(\min_{0\leq \tau \leq t} \xi_\tau \geq 0\right).
\end{aligned}
\end{equation}
The well-known results on the probability distribution of the first hitting time~\cite{patie2008first} allow us to obtain~\eqref{eq:safety_prob_hitting_time} as a solution to the two-dimensional PDE given by
\begin{equation}
\label{eq:one_dim_PDE_safe_prob}
\begin{aligned}
    W_{F}(\xi,t) := & \frac{\partial F}{\partial t}(\xi, t) - \alpha(\xi, t)\beta(\xi, t)\frac{\partial F}{\partial \xi}(\xi, t) \\
    & \quad - \frac{1}{2} \alpha(\xi, t) \frac{\partial^2 F}{\partial \xi^2}(\xi, t) =0,
\end{aligned}
\end{equation}
with boundary and initial conditions
\begin{equation}\label{eq:safe_PDE_ic_bc}
      F(\xi, 0)  = 1, \,\, \xi > 0; \quad F(0, t)  = 0, \,\, t > 0.
\end{equation}
  

\subsection{Physic-informed Learning}

In this section, we will introduce a physics-informed learning pipeline to solve the PDE for the value function~\eqref{eq:one_dim_PDE} and the PDE for safety probability~\eqref{eq:one_dim_PDE_safe_prob}. For conciseness, we will focus on the case of value function estimation, as adaptation to safety probability estimation is trivial where one just needs to replace the variables and the governing PDE.

From~\eqref{eq:path_integral_MC_estimation}, we can estimate the value function using path integral control by sampling the uncontrolled process. However, the path integral MC is not sample efficient when we want to know the value function on the entire space, especially for a large horizon $T$. Further, efficiently solving the value function PDE~\eqref{eq:one_dim_PDE} using standard numerical methods is challenging. 

\begin{figure*}[t]
\begin{subfigure}[t]{0.65\textwidth}
\centering
\input{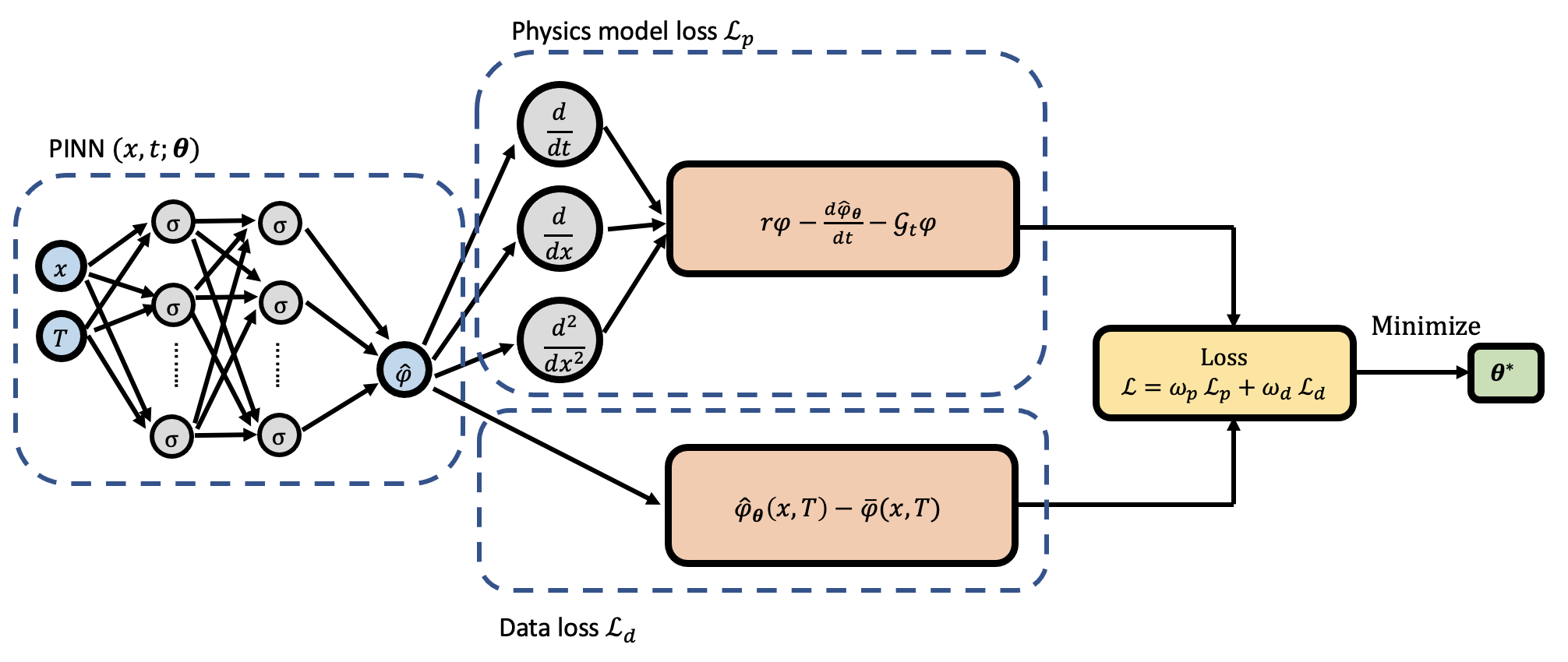}
\phantomsubcaption
\label{fig:PINN diagram}
\end{subfigure}
\hfill
\begin{subfigure}[t]{0.35\textwidth}
\centering
\phantomsubcaption
\input{Figures/AE_diagram}
\label{fig:AE diagram}
\end{subfigure}
\caption{(\subref{fig:PINN diagram}) The training scheme of the physics-informed neural network (PINN).
        (\subref{fig:AE diagram}) Autoencoder-like network architecture.}
\end{figure*}

To leverage the advantages of MC and PDE methods and to overcome their drawbacks, we propose a physics-informed neural network (PINN) to learn the mapping from the feature-time pair to the value function $\varphi$.
Fig.~\ref{fig:PINN diagram} shows the architecture of the PINN. The PINN takes the feature-time pair $(\xi,t)$ as the input, and outputs the value function prediction $\hat{\varphi}$, the feature and time derivatives $ \frac{\partial \hat \varphi}{\partial \xi}$ and $ \frac{\partial \hat \varphi}{\partial t}$, and the Hessian $ \frac{\partial^2 \hat \varphi}{\partial \xi^2}$, which come naturally from the automatic differentiation in deep learning frameworks such as PyTorch~\cite{paszke2019pytorch} and TensorFlow~\cite{abadi2016tensorflow}.
Assume the PINN is parameterized by $\boldsymbol\theta$, the loss function is defined as
\begin{equation}
\label{eq:PINN overall loss function}
    \mathcal{L}(\boldsymbol\theta) = \omega_p \mathcal{L}_p(\boldsymbol\theta) + \omega_d \mathcal{L}_d(\boldsymbol\theta),
\end{equation}
where
\begin{equation}
\label{eq:PINN loss functions}
\begin{aligned}
    \mathcal{L}_p(\boldsymbol\theta) & = \frac{1}{|\mathcal{P}|} \sum_{(\xi,t) \in \mathcal{P}} \|W_{\hat{\varphi}_{\boldsymbol\theta}}(\xi,t)\|_2^2, \\
    \mathcal{L}_d(\boldsymbol\theta) & = \frac{1}{|\mathcal{D}|} \sum_{(\xi,t) \in \mathcal{D}} \|\hat{\varphi}_{\boldsymbol\theta}(\xi,t) - \bar{\varphi}(\xi,t)\|_2^2.
\end{aligned}
\end{equation}
Here, $\bar{\varphi}$ is the training data, $\hat{\varphi}_{\boldsymbol\theta}$ is the prediction from the PINN, $\mathcal{P}$ and $\mathcal{D}$ are the training point sets for the physics model and external data, respectively. The loss function $\mathcal{L}$ consists of two parts, the physics model loss $\mathcal{L}_p$ and data loss $\mathcal{L}_d$. The physics model loss $\mathcal{L}_p$ measures the satisfaction of the PDE constraints for the learned output. It calculates the actual PDE equation value $W_{\hat{\varphi}_{\boldsymbol\theta}}$, which is supposed to be $0$, and use its 2-norm as the loss. The data loss $\mathcal{L}_d$ measures the accuracy of the prediction of PINN on the training data. It calculates the mean square error between the PINN prediction and the training data point as the loss. The overall loss function $\mathcal{L}$ is the weighted sum of the physics model loss and data loss with weighting coefficients $\omega_p$ and $\omega_d$.
Though out of the scope of this paper, theoretical analysis on the approximation error of PINNs and neural operators can be found in~\cite{wang2023generalizable, lu2021learning, kovachki2021universal}. 

\begin{remark}
For finding the optimal control, one can further enhance the sample complexity using path integral control together with value function prediction from the PINN. Denote $\hat{V}$ as the PINN value function prediction, we can initially estimate the optimal control from~\eqref{eq:optimal_control_from_V} with $\hat{u}(x,t) =  -\sigma\left(x\right)^\top \nabla \hat{V}\left(x, t\right)$. Then we can refine the optimal control using importance sampling~\cite{thijssen2015path} with the following procedure
\begin{equation}
\label{eq:optimal_control_IS}
\begin{aligned}
& u^{\star}(x, t)-\hat{u}(x, t)= \\
& \lim _{s \searrow t} \frac{\mathbb{E}^{P}\left\{\exp \left\{-S^{\hat{u}}(t)\right\} \int_t^s d W_\tau \mid x_t=x\right\}}{(s-t) \mathbb{E}^{P}\left\{\exp \left\{-S^{\hat{u}}(t)\right\} \mid x_t=x\right\}}
\end{aligned}
\end{equation}
where $P$ is the process with regard to $\hat{u}$ and
\begin{equation}
\label{eq:cost_function_u_hat}
    S^{\hat{u}}(t)=\int_t^T w\left(x_\tau, \hat{u}_\tau\right)
    d \tau+\int_t^T \hat{u}_\tau^{\top} d W_\tau+c\left(x_T\right).
\end{equation}
Essentially, one can use the sampled cost function~\eqref{eq:cost_function_u_hat} with control policy $\hat{u}$ to estimate the optimal control policy $u^\star$ with~\eqref{eq:optimal_control_IS}. The sample complexity for estimating the expectation in~\eqref{eq:optimal_control_IS} is much lower than the naive estimation with~\eqref{eq:path_integral_MC_estimation} and~\eqref{eq:optimal_control_from_V} due to the fact that path generated from $\hat{u}$ is much closer to the optimal path. The importance sampling theory provides theoretical analysis on the improvement of the sample complexity~\cite{thijssen2015path}.
    
\end{remark}

\subsection{Generalization: Arbitrary Feature Dimension}
In this section we generalize the previous results such that the representation of the system can be of arbitrary dimension. The procedure consists of two key steps. First, we use comparison theorem to find a multidimensional representation of the value function and the associated multidimensional process, \ie $\xi = [\xi^{(1)}, \xi^{(2)}, \cdots, \xi^{(k)}]^\top$ where $k$ is the dimension of the reduced representation. Then, we apply the high-dimensional Feynman-Kac formula~\cite[Theorem 1.3.17]{pham2009continuous} to
transform the stochastic process $\xi$ to a $k$-dimensional PDE which can be solved by the PINN.

For the first step, we find functions $p(x) = [p_1(x), p_2(x), \cdots, p_k(x)]^\top$ as the low dimensional representation of the original system. 
We define $a_i^{+-}$ and $b_i^{+-}$ similar to~\eqref{eq:a_b_def} and assume Assumption~\ref{asm:match_upper_lower_bound} and~\ref{asm:continuity-condition-on-a-and-b} hold for $\forall i$.
Then from the comparison theorem, we can find stochastic processes
\begin{equation}
\label{eq:xi_process_high_dim}
    d \xi^{(i)}_t=\alpha_i\left(\xi^{(i)}_t\right) \beta_i\left(\xi^{(i)}_t\right) d t+\sqrt{\alpha_i\left(\xi^{(i)}_t\right)} d \tilde{B}^{(i)}_t
\end{equation}
for $i = 1,2,\cdots, k$ that characterize $p_i(x)$, where $\tilde{B}^{(i)}_t$ is one-dimensional standard Wiener processes.

\begin{assumption}
\label{asm:value_function_representation}
    For value function estimation, we assume the running-cost can be represented by the following function
\begin{equation}
\label{eq:cost_function_xi}
    c(x) = r(\xi) = r\left(p_1(x), p_2(x), \cdots, p_k(x)\right)
\end{equation}
where $r:\mathbb{R}^{k} \rightarrow \mathbb{R}$ is a continuous function.
\end{assumption}
Assume Assumption~\ref{asm:value_function_representation} holds, then 
\begin{equation}
\label{eq:xi_value_function_high_dim}
\begin{aligned}
    \varphi(x, t) 
    & = \mathbb{E} \left[\exp \left(-\int_t^T r(\xi_\tau) d \tau - r(\xi_T)\right)  \mid \xi_t=\xi\right].
\end{aligned}
\end{equation}
From the high-dimensional Feynman-Kac formula~\cite[Theorem 1.3.17]{pham2009continuous}, we have $\varphi$ as the solution of the following PDE
\begin{equation}
\begin{aligned}
r \varphi -\frac{\partial \varphi}{\partial t}-\mathcal{G}_t \varphi =0&, & & \text { on } \mathbb{R}^k \times [0, T) \\
\varphi(\cdot,T) =\exp(-r(\cdot))&, & & \text { on } \mathbb{R}^k,
\end{aligned}
\end{equation}
where $r$ is given by~\eqref{eq:cost_function_xi} and
\begin{equation}
\label{eq:Gt_def}
   \mathcal{G}_t  (\cdot)=\alpha\left(\xi_t\right) \beta\left(\xi_t\right) \cdot \frac{\partial (\cdot)}{\partial \xi}
    +\frac{1}{2} \operatorname{Tr}\left(\mathbf{a}\left(\xi_t\right) \frac{\partial^2 (\cdot)}{\partial \xi^2}
    \right),
\end{equation}
with 
\begin{equation}
    \alpha\left(\xi_t\right) \beta\left(\xi_t\right) = \begin{bmatrix}
    \alpha_1\left(\xi^{(1)}_t\right) \beta_1\left(\xi^{(1)}_t\right) \\
    \alpha_2\left(\xi^{(2)}_t\right) \beta_2\left(\xi^{(2)}_t\right) \\
    \vdots \\
    \alpha_k\left(\xi^{(k)}_t\right) \beta_k\left(\xi^{(k)}_t\right)
    \end{bmatrix},
\end{equation}
and
\begin{equation}
    \mathbf{a}\left(\xi_t\right) = \begin{bmatrix}
    \alpha_1\left(\xi_t\right) &  &  \\
    & \alpha_2\left(\xi_t\right) &  &  \\
    &  & \ddots &  \\
    &  &  & \alpha_k\left(\xi_t\right) \\
    \end{bmatrix}.
\end{equation}

\begin{assumption}
\label{asm:barrier_function_representation}
Similarly, for safety probability estimation, we assume the barrier function can be represented by the following function
\begin{equation}
\label{eq:barrier_function_xi}
    \phi(x) = r(\xi) = r\left(p_1(x), p_2(x), \cdots, p_k(x)\right),
\end{equation}
where $r:\mathbb{R}^{k} \rightarrow \mathbb{R}$ is a continuous function.
\end{assumption}
Then the safety probability can be written as
\begin{equation}
\label{eq:xi_barrier_function_high_dim}
\begin{aligned}
    F(x,t) = \pr\left(\min_{0\leq \tau \leq t} \phi(x_\tau) \geq 0\right)  = \pr\left(\min_{0\leq \tau \leq t} r(\xi_\tau) \geq 0\right).
\end{aligned}
\end{equation}
We define $\mathcal{B} = \left\{\xi: r(\xi) \geq 0\right\}$. From the probability distribution of hitting time~\cite{patie2008first} we have that~\eqref{eq:xi_barrier_function_high_dim} can be characterized by the solution of the following PDE
\begin{equation}
    \frac{\partial F}{\partial t}-\mathcal{G}_t F =0,  \text {on }[0, T) \times \mathbb{R}^k
\end{equation}
\begin{equation}
   F(\xi,0)  =1, \,\, \xi\in\mathcal{B}; \quad F(\xi,t)  =0, \,\, \xi \in \partial \mathcal{B}.
\end{equation}


\begin{remark}
\label{rmk:low_dim_representation}
    The Assumptions~\ref{asm:value_function_representation} or~\ref{asm:barrier_function_representation} where the running-cost can be represented by~\eqref{eq:cost_function_xi} or the barrier function can be represented by~\eqref{eq:barrier_function_xi} is a necessary condition for the proposed method to work. The high-dimensional system must admit a low-dimensional representation of its value function/safety probability.
\end{remark}

\vspace{-0.4em}

\subsection{Deep Learning for Feature Identification}
For high dimensional systems with complex structure, deriving feature maps $p_1, p_2, \ldots, p_k$ such that Assumptions 2 and 3 hold is a challenging problem.  Thus, we propose an autoencoder-like neural network (Fig. \ref{fig:AE diagram}) to automatically identify lower-dimensional features that meet the bounding  requirements of Assumption~\ref{asm:match_upper_lower_bound} and sufficiently represent the cost/barrier function (Remark~\ref{rmk:low_dim_representation}). The network takes an input state $x\in\mathbb{R}^n$ and outputs a low-dimensional representation $\xi\in\mathbb{R}^k$ via the encoder $p_\sigma(x)$ and the function $\hat{r}(\xi)\in\mathbb{R}$ via the decoder $g_\sigma(\xi)$. We use $\boldsymbol \theta$ as the parameters of the autoencoder-like model. The loss function is defined as 
\begin{equation}\label{eq: AE Loss}
    \mathcal{L}_{AE}(\boldsymbol\theta) = w_{RC}\mathcal{L}_{RC}(\boldsymbol\theta) + w_{C.T.}\mathcal{L}_{C.T.}(\boldsymbol\theta)
\end{equation}
where
\begin{align} 
    \mathcal{L}_{RC}(\boldsymbol\theta) &= \frac{1}{|\mathcal{X}|}\sum_{x\in\mathcal{X}}\left(c(x)-\hat{r}(\xi; \boldsymbol\theta)\right)^2, \label{eq: RC loss} \\
    \mathcal{L}_{C.T.}(\boldsymbol\theta) &= \frac{1}{k}\sum_{i=1}^k\frac{1}{|\mathcal{R}_i|}\sum_{\xi\in\mathcal{R}_i}\frac{1}{|\mathcal{M}_{\xi, i}|}\sum_{x\in\mathcal{M}_{\xi, i}} \label{eq: CT loss} \\
    & \quad \ldots ||\nabla_x a_i(x; \boldsymbol\theta) ||_2^2 + || \nabla_x b_i(x; \boldsymbol\theta)||_2^2 \nonumber
\end{align}   
Here, $\mathcal{X}$ is the discretized state-space, $k$ is the dimension of the feature space, $\mathcal{R}_i$ is the range of $p_i$ and $\mathcal{M}_{\xi, i}=\{x: p_i(x; \boldsymbol\theta)=\xi\}$ is the set of $x$ that maps to the same value of $\xi\in\mathcal{R}_i$. The reconstruction loss $\mathcal{L}_{RC}$ measures how well the reconstruction $\hat{r}(\xi)$ represents $c(x)$ or $\phi(x)$, corresponding to Remark~\ref{rmk:low_dim_representation}. The comparison theorem loss $\mathcal{L}_{C.T.}$ enforces the condition that $a_i(x)$ and $b_i(x)$ are constant for all $x\in \mathcal{M}_{\xi, i}$, for each $\xi\in\mathcal{R}_i$, and for each feature $p_1,\ldots,p_k$. This is a sufficient condition to achieve Assumption~\ref{asm:match_upper_lower_bound}. The overall loss function is the weighted sum of the reconstruction loss and comparison theorem loss, where weights are chosen according to the desired strength and balance on the cost/barrier function reconstruction and the satisfaction of the comparison theorem. 
We refer readers to the appendix for more details about the algorithm. 

\vspace{-0.3em}

\section{Experiments}
\label{sec:experiments}

In this section, we show experiment results of the proposed method with both qualitative and quantitative analysis. 


\subsection{Value Function Estimation}

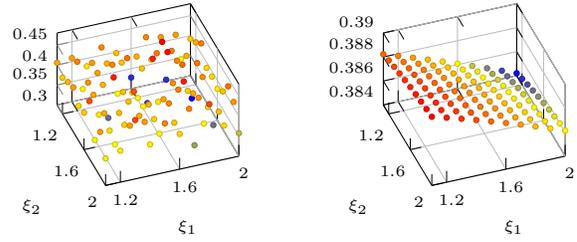
\begin{figure}[t]
  \begin{subfigure}{0.49\columnwidth}
  \input{Figures/path_integral_1kd.tex}
  \phantomsubcaption
  \end{subfigure}
  \begin{subfigure}{0.49\columnwidth}
  \input{Figures/prediction_1kd}
  \phantomsubcaption
  \end{subfigure} 
  \vspace{-0.5em}
  \caption{Estimation of the exponential of value function at $t=0.5$ for the 1000-dimensional system by path integral MC (left), and by the proposed method (right).}
  \label{fig:value_func_visual_1000d}
\end{figure}

We consider a 1000-dimensional system for which we want a 2-dimensional representation of the optimal value function. The system dynamics is given by
\begin{equation}
\label{eq:1000d_system_experiment}
\begin{aligned}
    dx = & \bar A x dt+ \sigma ( udt + dw), \\
\end{aligned}
\end{equation}
where $x \in \mathbb{R}^{1000}$ is the state, $u \in \mathbb{R}^{1000}$ in the control, and $dw$ is the 1000-dimensional standard Wiener process. We set $\sigma = I_{1000}$ to be the identity matrix, and set $\bar A = \left[\begin{array}{l|l}
A & 0 \\
\hline 0 & A
\end{array}\right]$ where $A \in \mathbb{R}^{500 \times 500}$. Let $a_{i,j}$ be the entry of $A$ at $i$-th row and $j$-th column. We choose $A$ such that $a_{i,i} = 1.1$, $a_{i,(i+2) | 500} = a_{i,(i+4) | 500} = 0.1$ and $a_{i,(i+6) | 500} = a_{i,(i+8) | 500} = -0.1$ for $\forall i = 1,2, \cdots, 500$, where $|$ is the mod operator. The running cost function is assumed to be
\begin{equation}
\label{eq:1000d_system_experiment_cost}
    c(x) = 
    \frac{1}{500}(\sum_{i=1}^{500} x_i)^2 + \frac{1}{500}(\sum_{j=501}^{1000}{x_j})^2.
\end{equation}
We pick two features of the state,
\begin{equation}
\label{eq:1000d_features}
    \xi_1 = p_1(x) = \sum_{i=1}^{500} x_i, \quad \xi_2 = p_2(x) = \sum_{j=501}^{1000} x_j.
\end{equation}
Then from~\eqref{eq:a_b_def}, we have
\vspace{-0.2em}
\begin{equation}
\label{eq:1000d_alpha_beta}
\begin{aligned}
    \alpha_1(\xi_1) = \alpha_2(\xi_2) = 500; \, \beta_1(\xi_1) = \frac{\xi_1}{500}, \ \beta_2(\xi_2) = \frac{\xi_2}{500},
\end{aligned}
\end{equation}
thus satisfying Assumptions~\ref{asm:match_upper_lower_bound} and~\ref{asm:continuity-condition-on-a-and-b}, and the running cost function can be written as
\begin{equation} \label{eq: r_xi_1000d}
    r(\xi) = \frac{1}{500}\xi_1^2 + \frac{1}{500}\xi_2^2,
\end{equation}
thus satisfying Assumption~\ref{asm:value_function_representation}. By the Feynman-Kac formula, we know that the exponential of the optimal value function $\varphi$ is the solution of the following PDE
\begin{align}
    & 0 = r(\xi)\mu 
    - \frac{\partial \mu}{\partial t} - 
    \alpha(\xi) \beta(\xi) \frac{\partial \mu}{\partial \xi} - \frac{1}{2} \operatorname{Tr} \left(
    \mathbf{a}(\xi) \frac{\partial^2 \mu}{\partial \xi^2} \right) \\
    & = \frac{\xi_1^2 + \xi_2^2}{500}\mu - \frac{\partial \mu}{\partial t} - \xi_1 \frac{\partial \mu}{\partial \xi_1} - \xi_2 \frac{\partial \mu}{\partial \xi_2} - \frac{\partial^2 \mu}{\partial \xi_1^2} - 250 \frac{\partial^2 \mu}{\partial \xi_2^2} \nonumber
\end{align}
\begin{equation}
    \mu(\xi, T) = \exp(- \frac{1}{500}\xi_1^2 - \frac{1}{500}\xi_2^2).
\end{equation}
With that, we generate data for $\bar \varphi(\xi,t)$ on spatial-temporal space $\Omega \times \mathbb{T} = [1,2]^2 \times [0,1.5]$, with grid size $d\xi = 0.1$ and $dt = 0.1$ and train the PINN on $\Omega \times \mathbb{T} = [1,2]^2 \times [1,1.5]$. 
We use a PINN with 3 hidden layers and 32 neurons per layer to learn the value function $\hat \varphi$. The activation function is chosen as hyperbolic tangent function ($\tanh$). We use the Adam optimizer~\cite{kingma2014adam} for training with initial learning rate set as $0.001$. The PINN parameters $\boldsymbol\theta$ are initialized via Glorot uniform initialization and the weights in the loss function~\eqref{eq:PINN overall loss function} are set to be $\omega_p = \omega_d = 1$. The simulation is constructed based on the DeepXDE framework~\cite{lu2021deepxde}. Fig.~\ref{fig:value_func_visual_1000d} shows the estimated value function from the path integral MC and the proposed method. It can be seen that the proposed framework is able to estimate value functions accurately, while the path integral method has significantly more noise. Also, the computation time for training the PINN is significantly less than sampling path integral MC (80s v.s $\sim$3000s). Note that the PINN is able to estimate the value function at unseen regions in the state space and generalize to longer time horizons, as the testing data at $t = 0.5$ is not seen by the PINN during training. 
We refer readers to the appendix for the safety probability estimation setting where similar results can be obtained.

\vspace{-0.1em}

\subsection{Sample Complexity}
We further examine the computation complexity of the proposed method to show its advantages in sample efficiency. We consider the problem of value function estimation of a 3-dimensional system with similar dynamics and cost defined in~\eqref{eq:1000d_system_experiment} and~\eqref{eq:1000d_system_experiment_cost}. 
Please see appendix for details of the settings.
Fig.~\ref{fig:sample_complexity} shows the percentage error of the estimated value function with different number of samples for path integral MC and PINN with and without using comparison theorem for dimension reduction. The path integral MC with dimension reduction has sample complexities that are one order less than MC without dimension reduction, which indicates the efficacy of the dimension reduction scheme from the comparison theorem. The PINN further reduces sample complexity and achieves the best trade-off between accuracy and computation.

\begin{figure}
    \centering
    \input{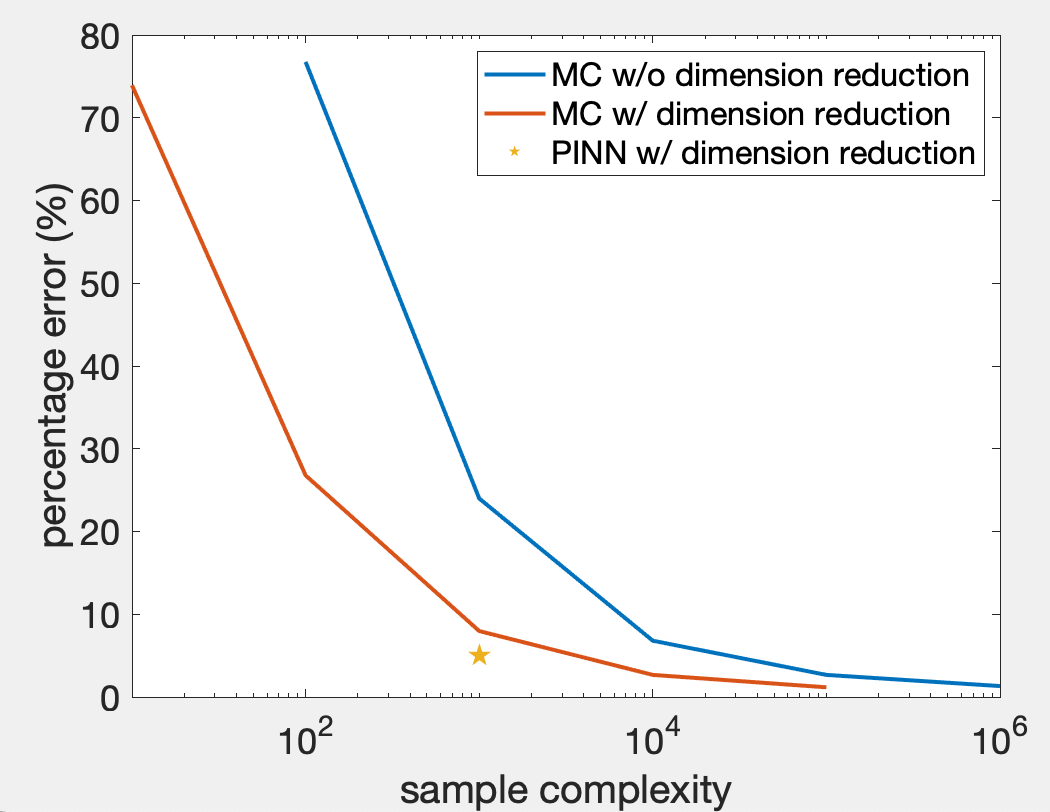}
    \vspace{-0.8em}
    \caption{Percentage error of the estimated value function with path integral MC.}
    \vspace{-0.5em}
    \resolved{refine the plot, use pdf format}
    \label{fig:sample_complexity}
\end{figure}

\vspace{-0.2em}

\subsection{Feature Learning}
\vspace{-0.05em}
We verify our autoencoder-like network by learning the features and cost function 
(similar to equations \eqref{eq:1000d_features} and~\eqref{eq: r_xi_1000d})
for the 3-dimensional system used for sample complexity analysis. The network consists of 5 fully connected hidden layers of sizes 100, 10, 2, 10, 100, respectively, with the activation function as hyperbolic tangent. We use the Adam optimizer with initial learning rate set as 0.001. The parameters of the network are initialized via the Glorot uniform initialization. The weights in the loss function \eqref{eq: AE Loss} are set as $w_{RC}=1, \,w_{C.T.}=10$. The network is trained on a $[0, 1]^3$ state-space with grid size 0.01. The network successfully identified the features derived analytically with $\mathrm{MSE}(\xi_1, \hat{\xi}_1)=0.2$ and $\mathrm{MSE}(\xi_2, \hat{\xi}_2)=0.06$. 

\vspace{-0.3em}

\section{Conclusion}
\label{sec:conclusion}

We propose a unified framework for value function and safety probability estimation of high-dimensional stochastic systems. The novel dimensionality reduction technique uses the comparison theorem to generate low-dimensional stochastic processes that provide an exact characterization of the cost/barrier function, significantly improving sample complexity. We then transform the low-dimensional process into a low-dimensional PDE, and leverage physics-informed learning to generalize solutions into longer time horizons and unseen regions of state-space. We also achieve automatic feature identification through a specially designed autoencoder-like neural network. Experiment results show the efficacy of the proposed method. Future work includes application to multi-agent robotic control systems.



\section{Acknowledgments}
This project is funded in part by Carnegie Mellon University’s Mobility21 National University Transportation Center, which is sponsored by the US Department of Transportation, in part by JST, PRESTO Grant Number JPMJPR2136, Japan, in part by the Department of the Navy, Office of Naval Research, grant number N00014-23-1-2252, in part by ACT-X, Japan, under Grant JPMJAX210L, and in part by AFOSR Grant FA9550-20-1-0101. Any opinions, findings, and conclusions or recommendations expressed in this material are those of the author(s) and do not necessarily reflect the views of the Office of Naval Research.

\bibliography{main}


\appendix

\section{Appendix}

\section{Proof of Theorems}
In this section, we provide proof for Theorem~\ref{thm:comparison_lemma}, which is based on~\cite[Theorem 3.1]{ikeda1977comparison}.

\begin{proof}(Theorem~\ref{thm:comparison_lemma})
    Let $\xi_t$ be a stochastic process that characterizes the evolution of $p(x_t)$ with $x_t$ being sampled from system~\eqref{eq:continuous_dynamics} with $\xi_0 = p(x_0)$. From~\cite[Theorem 3.1]{ikeda1977comparison} we know the existence of such process $\xi_t$.
    Based on the definitions~\eqref{eq:a_b_def}, and according to~\cite[Theorem 3.1]{ikeda1977comparison}, we can construct the following four stochastic processes based on $a^\pm$ and $b^\pm$.
    \begin{equation}
    \label{eq:xi_process_++}
    \left\{\begin{aligned}
d \xi_t^{++} & =\sqrt{a^{+}\left(\xi_t^{++}\right)} d \widetilde{B}_t^{+}+a^{+}\left(\xi_t^{++}\right) b^{+}\left(\xi_t^{++}\right) d t, \\
\xi_0^{++} & =\xi_0 ,
\end{aligned}\right.
    \end{equation}
    \begin{equation}
    \label{eq:xi_process_+-}
        \left\{\begin{aligned}
d \xi_t^{+-} & =\sqrt{a^{+}\left(\xi_t^{+-}\right)} d \widetilde{B}_t^{+}+a^{+}\left(\xi_t^{+-}\right) b^{-}\left(\xi_t^{+-}\right) d t, \\
\xi_0^{+-} & =\xi_0,
\end{aligned}\right.
    \end{equation}
    \begin{equation}
    \label{eq:xi_process_-+}
        \left\{\begin{aligned}
d \xi_t^{-+} & =\sqrt{a^{-}\left(\xi_t^{-+}\right)} d \tilde{B}_t^{-}+a^{-}\left(\xi_t^{-+}\right) b^{-}\left(\xi_t^{-+}\right) d t, \\
\xi_t^{-+} & =\xi_0,
\end{aligned}\right.
    \end{equation}
    \begin{equation}
    \label{eq:xi_process_--}
        \left\{\begin{aligned}
d \xi_t^{--} & =\sqrt{a^{-}\left(\xi_t^{--}\right)} d \tilde{B}_t^{-}+a^{-}\left(\xi_t^{--}\right) b^{-}\left(\xi_t^{--}\right) d t, \\
\xi_0^{--} & =\xi_0,
\end{aligned}\right.
    \end{equation}
    where both $\widetilde{B}_t^{+}$ and $\widetilde{B}_t^{-}$ are one-dimensional Wiener processes, and we have $\xi_t^{--} \leq  \xi_t \leq \xi_t^{++}$ with probability 1.
    
    Given Assumption~\ref{asm:match_upper_lower_bound} holds, we know that $a^+(\xi) = a^-(\xi) = \alpha(\xi)$ and $b^+(\xi) = b^-(\xi) = \beta(\xi)$, $\forall \xi \in I$. Thus, the four stochastic processes~\eqref{eq:xi_process_++} - \eqref{eq:xi_process_--} are identical, which gives 
    \begin{equation}
        \xi_t^{++} = \xi_t^{+-} = \xi_t^{-+} = \xi_t^{--} = \xi_t.
    \end{equation}
    Replacing $\xi_t^{\pm \pm}$ with $\xi_t$, $a^\pm$ with $\alpha$ and $b^\pm$ with $\beta$, we have that $p(x_t)$ with $x_t$ being sampled from system~\eqref{eq:continuous_dynamics} is characterized by the following stochastic process
    \begin{equation}
    d \xi_t = \alpha\left(\xi_t\right) \beta\left(\xi_t\right) d t + \sqrt{\alpha\left(\xi_t\right)} d \tilde{B}_t,
\end{equation}
with $\xi_0 = p(x_0)$, and $\tilde{B}_t$ being a one-dimensional standard Wiener process.

\end{proof}

\section{Details of the Sample Complexity Analysis}

In this section, we provide details of the sample complexity analysis. 

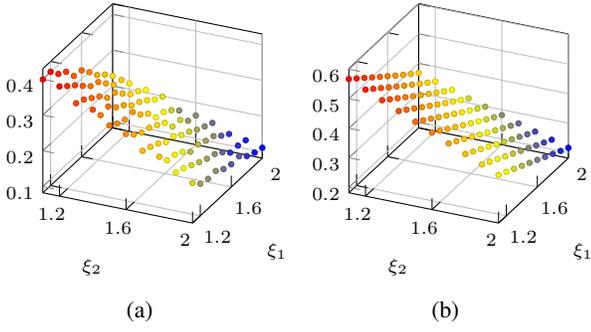
\begin{figure}[t]
  \begin{subfigure}{0.45\columnwidth}
  \input{Figures/value_func_visual_a}
  \caption{}
  \end{subfigure}
  \hspace{3pt}
  \begin{subfigure}{0.45\columnwidth}
  \input{Figures/value_func_visual_b}
  \caption{} 
  \end{subfigure} 
  \caption{Exponential of value function at $t=0.5$ (a) from solving the optimal control with Riccati equations, (b) from the PINN prediction.}
  \label{fig:value_func_visual}
\end{figure}

We consider a 3-dimensional system in which we want a 2-dimensional representation of the optimal value function. The system dynamics is given by
\begin{equation}
\label{eq:3d_system_experiment}
\begin{aligned}
    dx_1 &= (x_1 + x_3) \; dt + \sigma_1 (u_1 dt + dW_1) \\
    dx_2 &= (x_2 - x_3) \; dt + \sigma_2 (u_2 + dW_2) \\
    dx_3 &= x_3 \; dt + \sigma_3 (u_3 + dW_3)
\end{aligned}
\end{equation}
where $dW_i$ for $i = 1,2,3$ are standard Brownian motions and we assume $\sigma_i = 1$ for $i = 1,2,3$.
The running cost function is assumed to be
\begin{equation}
\label{eq:3d_system_experiment_cost}
    c(x) = \frac{1}{2}(x_1+x_2)^2 + \frac{1}{2}x_3^2.
\end{equation}
We pick two features of the state,
\begin{equation}
\label{eq:3d_features}
    \xi_1 = p_1(x) = x_1+x_2, \quad \xi_2 = p_2(x) = x_3.
\end{equation}
Then from~\eqref{eq:a_b_def}, we have
\begin{equation}
\label{eq:3d_alpha_beta}
\begin{aligned}
    \alpha_1(\xi_1) = 2, \,\, \alpha_2(\xi_2) = 1; \quad \beta_1(\xi_1) = \frac{\xi_1}{2}, \,\, \beta_2(\xi_2) = \xi_2
\end{aligned}
\end{equation}
thus satisfying Assumptions 2 and 3.
The cost function can then be written as
\begin{equation} \label{eq: r_xi}
    r(\xi) = \frac{1}{2}\xi_1^2 + \frac{1}{2}\xi_2^2.
\end{equation}
By the Feynman-Kac formula, we know that the exponential of the optimal value function $\varphi$ is the solution of the following PDE
\begin{align}
    & 0 = r(\xi)\mu 
    - \frac{\partial \mu}{\partial t} - 
    \alpha(\xi) \beta(\xi) \frac{\partial \mu}{\partial \xi} - \frac{1}{2} \operatorname{Tr} \left(
    \mathbf{a}(\xi) \frac{\partial^2 \mu}{\partial \xi^2} \right) \\
    & = (\frac{1}{2}\xi_1^2 + \frac{1}{2}\xi_2^2)\mu - \frac{\partial \mu}{\partial t} - \xi_1 \frac{\partial \mu}{\partial \xi_1} - \xi_2 \frac{\partial \mu}{\partial \xi_2} - \frac{\partial^2 \mu}{\partial \xi_1^2} - \frac{1}{2} \frac{\partial^2 \mu}{\partial \xi_2^2} \nonumber
\end{align}

\begin{equation}
    \mu(\xi, T) = \exp(- \frac{1}{2}\xi_1^2 - \frac{1}{2}\xi_2^2).
\end{equation}
With that, we generate data for $\bar \varphi(\xi,t)$ on feature-temporal space $\Omega \times \mathbb{T} = [1,2]^2 \times [0,1.5]$, with grid size $d\xi = 0.1$ and $dt = 0.1$ and train the PINN on $\Omega \times \mathbb{T} = [1,2]^2 \times [1,1.5]$. 
We use a PINN with 3 hidden layers and 32 neurons per layer to learn the value function $\hat \varphi$. The activation function is chosen as hyperbolic tangent function ($\tanh$). We use the Adam optimizer~\cite{kingma2014adam} for training with initial learning rate set as $0.001$. The PINN parameters $\boldsymbol\theta$ is initialized via Glorot uniform initialization and the weights in the loss function~\eqref{eq:PINN overall loss function} are set to be $\omega_p = \omega_d = 1$. Fig.~\ref{fig:value_func_visual} visualizes the estimated value function from the Riccati equation and the PINN prediction.

\section{Experiments for Safety Probability Estimation}

In this section, we present simulation results of the proposed method for safety probability estimation.

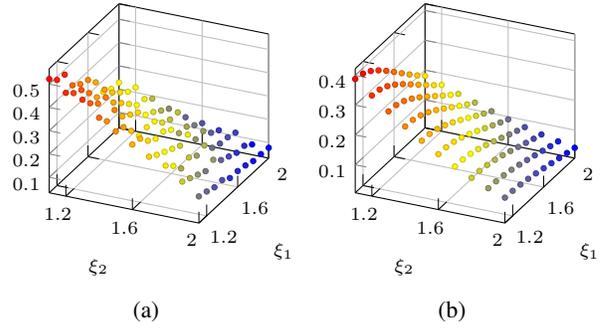
\begin{figure}[t]
  \begin{subfigure}{0.45\columnwidth}
  \input{Figures/safe_prob_visual_a}
  \caption{}
  \end{subfigure}
  \hspace{3pt}
  \begin{subfigure}{0.45\columnwidth}
  \input{Figures/safe_prob_visual_b}
  \caption{} 
  \end{subfigure} 
    \caption{Safety probability at $t = 1$ (a) from MC simulation, (b) from the PINN prediction.
    }
    \label{fig:safe_prob_3d->2d}
\end{figure}

We consider the same 3-dimensional system~\eqref{eq:3d_system_experiment}
and we set the nominal controller to be $\mathcal{N}(x) \equiv 0$. We consider safe set~\eqref{eq:safe set definition} with barrier function $\phi(x) = \min\{-(x_1+x_2), -x_3\}+4$. We pick the same features as~\eqref{eq:3d_features} and we have $\alpha_i$ and $\beta_i$ as in~\eqref{eq:3d_alpha_beta} for $i = 1,2$. Then the barrier function can be written as
\begin{equation}
    r(\xi) = \min\{-\xi_1, -\xi_2\}+4,
\end{equation}
and the set $\mathcal{B} = \{\xi: r(\xi) \geq 0\} = (-\infty,4] \times (-\infty,4]$. The PDE that governs the evolution of the safety probability is given by
\begin{equation}
\begin{aligned}
    & 0 = \frac{\partial \mu}{\partial t} - 
    \alpha(\xi) \beta(\xi) \frac{\partial \mu}{\partial \xi} - \frac{1}{2} \operatorname{Tr} \left(\mathbf{a}(\xi) \frac{\partial^2 \mu}{\partial \xi^2} \right)  \\
    & = \frac{\partial \mu}{\partial t} - \xi_1 \frac{\partial \mu}{\partial \xi_1} - \xi_2 \frac{\partial \mu}{\partial \xi_2} - \frac{\partial^2 \mu}{\partial \xi_1^2} - \frac{1}{2} \frac{\partial^2 \mu}{\partial \xi_2^2} 
\end{aligned}
\end{equation}
\begin{equation}
    \mu(\xi,0)  =1, \,\, \xi \in \mathcal{B}; \quad \mu(\xi,t) =0, \,\, \xi \in \partial \mathcal{B}.
\end{equation}
We use the same procedure to train a PINN for $\bar F(\xi,t)$ on feature-temporal space $\Omega \times \mathbb{T} = [1,2]^2 \times [0,1]$, with grid size $d\xi = 0.1$ and $dt = 0.1$. Fig.~\ref{fig:safe_prob_3d->2d} shows the estimated safety probability from MC simulation and the PINN prediction. 
We can see that the proposed method is able to estimate safety probability at unseen regions in the state space, generalize to longer time horizons, and does so with higher accuracy than the MC method.

\section{Physics Informed Learning Ablation Experiments}

In this section, we provide ablation results for the experiments presented in the paper on physics-informed neural networks (PINNs). 

\subsection{PINN Prediction on Different Time Horizons}

\begin{figure}[ht]
    \centering
    \input{Figures/prediction_time}
    \caption{Prediction on different time horizon.}
    \label{fig:prediction_time}
\end{figure}
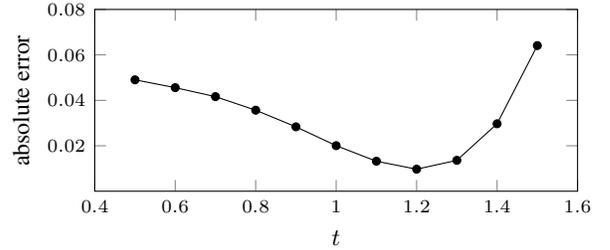

We record the absolute error of PINN predictions across different time horizons as shown in Fig.~\ref{fig:prediction_time}. The absolute errors remain consistently small (less than 0.08) from $t = 0.5$ to $t = 1.5$ throughout the entire time span, suggesting the model is able to generalize to unseen time horizons with high accuracy. 

\subsection{PINN Performance with Different Training Epochs}

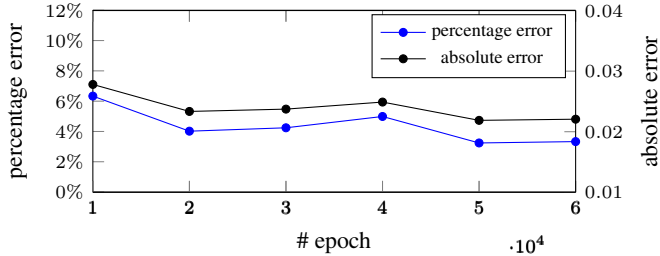
\begin{figure}[t]
    \centering
    \input{Figures/pinn_epoch}
    \vspace{-1.5em}
    \caption{PINN with different epochs for training (\#domain = 600).}
    \label{fig:pinn_epoch}
\end{figure}
\input{Tables/pinn_epoch_table}

Here, we assess the performance of the PINN across the amount of training epochs. We present the absolute error (absolute value of the difference between the PINN prediction and the ground truth) and the percentage error (absolute error over the ground truth value). Fig.~\ref{fig:pinn_epoch} and Table~\ref{tab:pinn_epoch} shows the results. 
The results suggest that PINN with long enough training epochs will yield accurate estimations, while further increasing training epochs (beyond  50000 epochs) will not give performance improvement.



\subsection{PINN Performance on Different PDE Constraints}

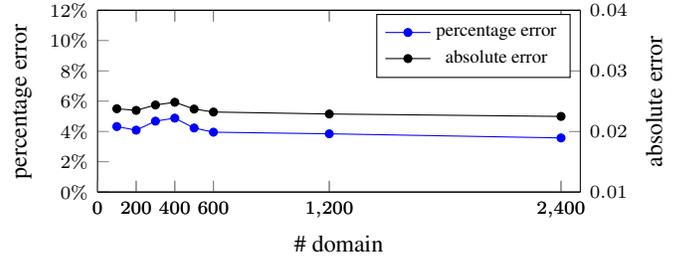
\begin{figure}[ht]
    \centering
    \input{Figures/pinn_pde}
    \vspace{-0.2em}
    \caption{PINN with different numbers of PDE constraint grid points.}
    \label{fig:pinn_pde}
\end{figure}

\input{Tables/pinn_pde_table}

\input{Tables/ae_table1}
\input{Tables/ae_table2}
\input{Tables/ae_table3}

We also examine the performance of the PINN with different PDE loss configurations. We fix the training epochs to be 60000 and vary the number of grid points for imposing PDE constraints for PINN training. We present the results in Fig.~\ref{fig:pinn_pde} with the numerical results shown in Table~\ref{tab:pinn_pde}. 
We can see the error drops as the PDE constraint is imposed with finer grids, but quickly saturates. For optimal training efficiency, we suggest to select a moderate number of grid points for PDE constraints in the PINN to balance accuracy with computation time. 

\vspace{-0.2em}

\section{Representation Learning with Autoencoder-like Neural Networks}

In this section, we provide details on the experiments presented in the paper on the autoencoder-like neural network.

\subsection{Algorithm Details}
We present details of the training procedure for the proposed autoencoder-like neural network (Fig. \ref{fig:AE diagram}) to automatically identify lower-dimensional features in Algorithm~\ref{alg:auto-encoder_training}. We use boldface symbols to denote the vector of such symbols (\eg $\boldsymbol\xi = [\boldsymbol\xi^{(1)}, \boldsymbol\xi^{(2)}, \ldots, \boldsymbol\xi^{(k)}]$, and $\boldsymbol\xi^{(1)}=[\xi_1^{(1)}, \xi_2^{(1)}, \ldots]^\top$). 

\begin{algorithm*}[t]
\caption{Autoencoder-like Neural Network Training}\label{alg:auto-encoder_training}
\begin{algorithmic}[1]
\State \textbf{Given:} $w_{RC}, w_{C.T.}, k, d, \alpha$ \Comment{$d$ is the frequency for preimage update, $\alpha$ is the initial learning rate}
\For {each epoch}
    \For {each batch}
        \State $\hat{r}(\boldsymbol\xi), \boldsymbol\xi \gets \textrm{AE}(\mathcal{X})$ \Comment{pass the batched state-space samples $\mathcal{X}$ to the autoencoder-like network AE to get $\hat{r}(\boldsymbol\xi), \boldsymbol\xi$} \label{line: AE pass}
        \If {$d$ batches have passed $\mathbf{or}$ first batch} \label{line:every_d_iter}
            \State $\mathcal{R}\gets\textrm{range}(\boldsymbol p)$ \Comment{find $\mathcal{R}$, the range of each $p_i$ for $i\in[k]$}
            \State 
            $\mathcal{M}_{\boldsymbol\xi} \gets \textrm{preimage}(\mathcal{X}, \boldsymbol\xi)$ \Comment{find $\mathcal{M}_{\boldsymbol\xi}$, the preimage of each $\boldsymbol\xi^{(i)}$ for $i\in[k]$} \label{line: preimage}
        \EndIf
        \State $\mathcal{L}_{RC}(\boldsymbol\theta) = \frac{1}{|\mathcal{X}|}\sum_{x\in\mathcal{X}}\left(c(x)-\hat{r}(\boldsymbol\xi; \boldsymbol\theta)\right)^2$ \Comment{running-cost/barrier function reconstruction loss} \label{line: RC loss}
        \State $ \mathcal{L}_{C.T.}(\boldsymbol\theta) = \frac{1}{k}\sum_{i=1}^k\frac{1}{|\mathcal{R}_i|}\sum_{\xi\in\mathcal{R}_i}\frac{1}{|\mathcal{M}_{\xi, i}|}$ \Comment{comparison theorem loss} \label{line: CT loss}
        \State \qquad\qquad\qquad\qquad
        $\ldots \sum_{x\in\mathcal{M}_{\xi, i}}
          ||\nabla_x a_i(x; \boldsymbol\theta) ||_2^2 + || \nabla_x b_i(x; \boldsymbol\theta)||_2^2$
        \State $\mathcal{L}_{AE}(\boldsymbol\theta)=w_{RC}\mathcal{L}_{RC}(\boldsymbol\theta)+w_{C.T.}\mathcal{L}_{C.T.}(\boldsymbol\theta)$ \Comment{total loss for the autoencoder} \label{line: total loss}
        \State $\boldsymbol\theta\gets\textrm{Adam}(\boldsymbol\theta, \alpha)$ \Comment{parameters update with the Adam optimizer}
    \EndFor
\EndFor
    
\end{algorithmic}
\end{algorithm*}

First, we partition the state-space training data into appropriately sized batches, $\mathcal{X}$. This batch is then passed into the autoencoder (line~\ref{line: AE pass}), where the encoder returns $\hat{\boldsymbol\xi}$ and the decoder returns $\hat{r}(\boldsymbol\xi)$. In order to obtain the comparison theorem loss in~\eqref{eq: CT loss}, we must calculate gradients over the preimage of each $\xi\in\mathcal{R}_i$.
Retrieving this set requires a method for inverting the encoder $p_i(x)$, and we approximate the preimage using a grid-search over the state-space (line~\ref{line: preimage}).
However, such a search can impose a high computational burden on the training regime, and to address this issue we update the preimage only after every $d$ batches (line~\ref{line:every_d_iter}). 
Thus, the comparison theorem loss in line~\ref{line: CT loss} uses the same (outdated) preimage for the following $d$ iterations. In the simulation described in the paper, we set $d=1$ since the preimage computation is not very expensive for the 3-dimensional system. However, for larger systems, it will become necessary to update the preimage sparsely to avoid the grid-search (or an alternative polyhedral computation~\cite{polyhedral}) over the high-dimensional state-space. 

Due to the discretization in $\mathcal{X}$, the preimage function in line~\ref{line: preimage} assumes a threshold $\epsilon$ for which 
$$|p_j(x_a) - p_j(x_b)|<\epsilon\implies \xi_a^{(j)}=\xi_b^{(j)},$$
where $a$ and $b$ are distinct samples of the state-space and $p_j(x)$ is the $j^{th}$ feature map of $k$ such maps. We choose $\epsilon$ empirically such that the encoder $p(x)$ is a many-to-one mapping, \ie $p_j: \mathcal{X}\in\mathbb{R}^{N\times n}\mapsto\boldsymbol\xi^{(j)}\in\mathbb{R}^{M\times 1}$, where $N$ is the number of state-space observations in $\mathcal{X}$, $M$ is the number of distinct feature observations $\boldsymbol{\xi}^{(j)}$ according to the threshold $\epsilon$, and $M\ll N$. In 3-dimensional example considered in the paper ($n=3$), $\epsilon_j=\frac{1}{5}\frac{1}{N-1}\sum_{i=1}^{N-1}|\xi_{i+1}^{(j)}-\xi_i^{(j)}|$, which results in $M\approx 15$ for $N=1000$.

In line~\ref{line: RC loss}, we calculate the reconstruction loss~\eqref{eq: RC loss} that quantifies how well the learned features $\boldsymbol\xi$ represent the cost function. The total loss in line~\ref{line: total loss} is the weighted sum of the reconstruction loss and the comparison theorem loss. The weights $w_{C.T.}$ and $w_{RC}$ should be chosen to ensure that $w_{C.T.}\mathcal{L}_{C.T.}$ and $w_{RC}\mathcal{L}_{RC}$ are similar in magnitude. In our test case, the state-space domain is $[0, 1]^3$
sampled at a spatial resolution of 0.01. Due to the smoothness feature of the encoder, the preimage $\mathcal{M}_{\xi, i}$ of any individual feature observation will have state-space observations contained in a small neighborhood whose density and width is determined by the sampling resolution and the threshold $\epsilon$, respectively. 
The Jacobians and Hessians needed to calculate $a(x)$ and $b(x)$ over this neighborhood could be small in magnitude, resulting in smaller $\mathcal{L}_{C.T.}$ relative to $\mathcal{L}_{RC}$. Thus, we choose $w_{C.T.}=1$ and $w_{RC}=10$ for our test case.  


\subsection{Network Performance on Different Training Regimes}
In this section, we compare the representation learning results for different network architectures in the proposed autoencoder-like model to examine its effect on performance. The network settings we consider are different hidden layer numbers and different numbers of neurons per layer. Recall the the basic model architecture contains two components: the encoder subarchictecture and the encoder subarchictecture. The encoder architecture consists of an projection from the system dimension to a high-dimensional space, followed by a projection down to the desired feature dimension. The decoder architecture is identical, only the down-projection is to a scalar value that represents the networks prediction of the cost/barrier function. Thus, to configure the network, we consider a different number of layers and neurons per layer in the up-projection and down-projection regions separately. For all experiments, we train the network on a $[0, 1]^3$ state-space with grid-size 0.01 using the Adam optimizer~\cite{kingma2017adam} with initial learning rate 0.001 and Glorot uniform initialization for the network parameters. The network is trained for 5 epochs with batch size of 1000 for 100 iterations. 

To examine the effect of the network architecture on feature learning, we begin with the encoder architecture and fix the number of layers in the decoder architecture at two, with 10 and 100 neurons respectively. We consider the number of layers in the encoder architecture as $n_{\text{layer}}\in[1, 3, 5]$. For each layer configuration, we consider the number of neurons per layer as $n_{\text{unit}}\in[16, 64, 128]$. Tables~\ref{tab:n_layer = 1}--\ref{tab:n_layer = 5} report the percent error of the learned features for different layer numbers and neurons per layer. It is clear that a wider network (more layers) learns features with error orders of magnitude lower than the shallow network experiments, yet sacrifices in computation time.

To examine the effect of the network architecture on cost function reconstruction, we fix the number of layers in the encoder architecture at two, with 10 and 100 neurons respectively. Experiments on the decoder architecture yield similar results (Tables~\ref{tab:n_layer = 1}--\ref{tab:n_layer = 5}). Increasing the number of neurons per layer had little effect on the ability of the model to reconstruct the cost function. However, wider networks consistently achieved lower percent error between the ground truth cost function and the predicted cost function. 

Thus, we find that the performance of the network quickly saturates at a relatively low number of units per layer. Best performance is observed when the encoder and decoder architecture are symmetric, with a significant gain in prediction accuracy in wider networks. 



\end{document}

%% file: Figures/overall_diagram.tex

\definecolor{color5}{HTML}{F2CC8F}
\definecolor{color2}{HTML}{F4F1DE}
\definecolor{color3}{HTML}{3D405B}
\definecolor{color4}{HTML}{81B29A}
\definecolor{color1}{HTML}{E07A5F}
\tikzset{
diagonal fill/.style 2 args={fill=#2, path picture={
\fill[#1, sharp corners] (path picture bounding box.south west) -|
                         (path picture bounding box.north east) -- cycle;}},
reversed diagonal fill/.style 2 args={fill=#2, path picture={
\fill[#1, sharp corners] (path picture bounding box.north west) |- 
                         (path picture bounding box.south east) -- cycle;}}
}

\begin{tikzpicture}[
    node distance = 4cm,
    node/.style={circle, fill=#1, inner sep=2pt, minimum size=0.7cm}, 
    edge/.style={-stealth, #1}
]

\node[rectangle, draw=cmuyellow, dashed, line width=0.5mm, minimum height=0.8cm,minimum width=2.2cm,inner sep=-3pt] (t1){\scriptsize \begin{tabular}{ll}
Inference of \\ cost and risk
\end{tabular}};
\node[shading = axis, rectangle, left color= cmuyellow!70, right color= cmublue!70, minimum height=0.8cm,minimum width=1.5cm, rounded corners,inner sep=0pt] (control)[right=0.3cm of t1] {\scriptsize \begin{tabular}{cc}
Control \\ Prospective
\end{tabular}};

\node[rectangle, draw=cmublue, dashed, line width=0.5mm, minimum height=0.8cm,minimum width=2.2cm,inner sep=0pt] (t2) [right=0.3cm of control]{\scriptsize \begin{tabular}{ll}
Reduction \\ in space
\end{tabular}};

\node[shading = axis, rectangle, top color= cmuyellow!70, bottom color= orange!70, minimum height=0.8cm,minimum width=1.5cm, rounded corners, inner sep = 0pt] (sr) [below=0.3cm of t1] {\scriptsize \begin{tabular}{cc}
Data \\ \& Sampling\\
\end{tabular}};

\begin{scope}[on background layer]
\node[ellipse, draw = cmured, line width=0.5mm, minimum width = 3cm, minimum height = 2cm, rounded corners] (pm)[right= -0.1 cm of sr]{\footnotesize \textcolor{cmured}{\begin{tabular}{cc}
Proposed \\Method
\end{tabular}}};
\end{scope}

\node[shading = axis, rectangle, top color= cmublue!70, bottom color= cmutan!70, minimum height=0.8cm,minimum width=1.5cm, rounded corners] (pde) [below=0.3cm of t2] {\scriptsize PDE};

\node[rectangle, draw=orange, dashed, line width=0.5mm, minimum height=0.8cm,minimum width=2.2cm,inner sep=0pt] (t3)[below=0.3cm of sr]{\scriptsize \begin{tabular}{cc}
Improved \\ sample complexity
\end{tabular}};

\node[shading = axis, rectangle, left color= orange!70, right color= cmutan!70,  minimum height=0.8cm,minimum width=1.5cm, rounded corners] (pinn)[right=0.3cm of t3]{\scriptsize PINN};

\node[rectangle, draw=cmutan, dashed, line width=0.5mm, minimum height=0.8cm,minimum width=2.2cm] (pw)[below=0.3cm of pde,inner sep=0pt]{\scriptsize \begin{tabular}{ll}
Reduction in \\ time horizon 
\end{tabular}};

\draw[edge=cmuyellow,line width=0.5mm] (control) -> (t1);
\draw[edge=cmublue,line width=0.5mm] (control) -> (t2);
\draw[edge=cmuyellow,line width=0.5mm] (sr) -> (t1);
\draw[edge=cmublue,line width=0.5mm] (pde) -> (t2);
\draw[edge=orange,line width=0.5mm] (pinn) -> (t3);
\draw[edge=cmutan,line width=0.5mm] (pde) -> (pw);
\draw[edge=cmutan,line width=0.5mm] (pinn) -> (pw);
\draw[edge=orange,line width=0.5mm] (sr) -> (t3);
\end{tikzpicture}

%% file: Figures/flowchart.tex
\definecolor{color5}{HTML}{F2CC8F}
\definecolor{color2}{HTML}{F4F1DE}
\definecolor{color3}{HTML}{3D405B}
\definecolor{color4}{HTML}{81B29A}
\definecolor{color1}{HTML}{E07A5F}
\tikzset{
diagonal fill/.style 2 args={fill=#2, path picture={
\fill[#1, sharp corners] (path picture bounding box.south west) -|
                         (path picture bounding box.north east) -- cycle;}},
reversed diagonal fill/.style 2 args={fill=#2, path picture={
\fill[#1, sharp corners] (path picture bounding box.north west) |- 
                         (path picture bounding box.south east) -- cycle;}}
}

\begin{tikzpicture}[
    node distance = 4cm,
    node/.style={circle, fill=#1, inner sep=2pt, minimum size=0.7cm}, 
    edge/.style={-stealth, #1}
]

\node[rectangle, draw=black, fill = newblue,line width = 0.3mm, minimum height=0.8cm,minimum width=2cm,inner sep=-3pt] (hds){\scriptsize \begin{tabular}{cc}
High-dimension \\ System
\end{tabular}};

\node[rectangle, draw=black, fill = newblue,line width = 0.3mm, minimum height=0.8cm,minimum width=2cm,inner sep=-3pt][right = 3.5cm of hds] (ldf){\scriptsize \begin{tabular}{cc}
Low-dimension \\ Features
\end{tabular}};

\node[rectangle, draw=black, fill = newblue,line width = 0.3mm, minimum height=0.8cm,minimum width=2cm,inner sep=-3pt][below = 0.8cm of hds] (pde){\scriptsize \begin{tabular}{cc}
Low-dimension \\ PDE
\end{tabular}};

\node[rectangle, draw=black, fill = newblue,line width = 0.3mm, minimum height=0.8cm,minimum width=2cm,inner sep=-3pt][right = 3.5cm of pde] (vfsp){\scriptsize \begin{tabular}{cc}
Value Function \\ Safety Probability
\end{tabular}};

\draw[edge=black,line width = 0.5mm] (hds) -> (ldf);
\node[rectangle, line width = 0.3mm, minimum height=0.8cm,minimum width=1.8cm,inner sep=-3pt][right = 0.45cm of hds] (ct){\scriptsize \begin{tabular}{cc}
Comparison Theorem \\ Autoencoder-like NN
\end{tabular}};

\draw [edge=black,line width = 0.5mm, draw] (ldf.south) -- ++(0,-0.4) node(left, midway, pos=0.2){} -| (pde.north);

\node[rectangle][below = -0.08cm of ct] (fk){\scriptsize \begin{tabular}{cc}
\color{black}{Feynman-Kac Formula}
\end{tabular}};

\draw [edge=black,line width = 0.5mm] (pde) -- (vfsp) node[midway, above] {};


\node[rectangle, line width = 0.3mm, minimum height=0.3cm,minimum width=1cm,inner sep=-3pt][below = 0.35cm of fk] (pil){\scriptsize \begin{tabular}{cc}
Physics-informed Learning
\end{tabular}};

\end{tikzpicture}

%% file: Figures/PINN_diagram.tex
\tikzstyle{line} = [draw, -latex']

\tikzset{
diagonal fill/.style 2 args={fill=#2, path picture={
\fill[#1, sharp corners] (path picture bounding box.south west) -|
                         (path picture bounding box.north east) -- cycle;}},
reversed diagonal fill/.style 2 args={fill=#2, path picture={
\fill[#1, sharp corners] (path picture bounding box.north west) |- 
                         (path picture bounding box.south east) -- cycle;}}
}

\begin{tikzpicture}[
    node distance = 1.2cm,
    fontscale/.style = {font=\scriptsize},
    node/.style={circle, fill=#1, inner sep=0pt, minimum size=0.1cm}, 
    edge/.style={-stealth, #1},
    arr/.style = {semithick, -Stealth}
]
    
    \node[circle, fill = newblue, draw=black, line width=0.3mm, inner sep=0pt, minimum size=8pt]  (xi) {\scriptsize $\xi$};
    \node[circle, fill = newblue, draw=black, line width=0.3mm, inner sep=0pt, minimum size=8pt] (t) [below = 0.2cm of xi] {\scriptsize $t$};

    \node[circle, fill = newgray, draw=black, line width=0.3mm, inner sep=0pt, minimum size=8pt]  (sigma1)[above right = -0cm and 0.5cm of xi] {\scriptsize $\sigma$};
    \node[circle, fill = newgray, draw=black, line width=0.3mm, inner sep=0pt, minimum size=8pt]  (sigma2)[below = 0.2cm of sigma1] {\scriptsize $\sigma$};
    \node[circle, fill = newgray, draw=black, line width=0.3mm, inner sep=0pt, minimum size=8pt]  (sigma3)[below = 0.5cm of sigma2] {\scriptsize $\sigma$};

    \node[circle, fill = newgray, draw=black, line width=0.3mm, inner sep=0pt, minimum size=8pt]  (sigma4)[right = 0.3cm of sigma1] {\scriptsize $\sigma$};
    \node[circle, fill = newgray, draw=black, line width=0.3mm, inner sep=0pt, minimum size=8pt]  (sigma5)[right = 0.3cm of sigma2] {\scriptsize $\sigma$};
    \node[circle, fill = newgray, draw=black, line width=0.3mm, inner sep=0pt, minimum size=8pt]  (sigma6)[right = 0.3cm of sigma3] {\scriptsize $\sigma$};

    \node[circle, fill = newblue, draw=black, line width=0.3mm, inner sep=0pt, minimum size=8pt]  (phi)[below right = 0.1cm and 0.3cm of sigma5] {\scriptsize $\hat \varphi$ / $\hat F$};

    \draw[edge,line width=0.3mm] (xi) -> (sigma1);
    \draw[edge,line width=0.3mm] (xi) -> (sigma2);
    \draw[edge,line width=0.3mm] (xi) -> (sigma3);

    \draw[edge,line width=0.3mm] (t) -> (sigma1);
    \draw[edge,line width=0.3mm] (t) -> (sigma2);
    \draw[edge,line width=0.3mm] (t) -> (sigma3);

    \draw[edge,line width=0.3mm] (sigma1) -> (sigma4);
    \draw[edge,line width=0.3mm] (sigma1) -> (sigma5);
    \draw[edge,line width=0.3mm] (sigma1) -> (sigma6);

    \draw[edge,line width=0.3mm] (sigma2) -> (sigma4);
    \draw[edge,line width=0.3mm] (sigma2) -> (sigma5);
    \draw[edge,line width=0.3mm] (sigma2) -> (sigma6);

    \draw[edge,line width=0.3mm] (sigma3) -> (sigma4);
    \draw[edge,line width=0.3mm] (sigma3) -> (sigma5);
    \draw[edge,line width=0.3mm] (sigma3) -> (sigma6);

    \node at ($(sigma2)!.35!(sigma3)$) {\scriptsize \vdots};
    \node at ($(sigma5)!.35!(sigma6)$) {\scriptsize \vdots};

    \draw[edge,line width=0.3mm] (sigma4) -> (phi);
    \draw[edge,line width=0.3mm] (sigma5) -> (phi);
    \draw[edge,line width=0.3mm] (sigma6) -> (phi);

    \node[rectangle, draw=cmublue, fit=(xi) (sigma1) (sigma3) (phi), rounded corners, dashed, line width=0.3mm, inner sep=1mm] (block1) {};
    \node[above right] at (block1.north west){\scriptsize PINN $(\xi,t;\boldsymbol\theta)$};

    \node[circle, fill = newgray, draw=black, line width=0.3mm, inner sep=0pt, minimum size=18pt]  (par1)[above right= 1cm and 0.5cm of phi] {\scriptsize $\frac{\partial}{\partial\xi}$};
    \node[circle, fill = newgray, draw=black, line width=0.3mm, inner sep=0pt, minimum size=18pt]  (par2)[below= 0.2cm of par1] {\scriptsize $\frac{\partial^2}{\partial\xi^2}$};
PF    \node[circle, fill = newgray, draw=black, line width=0.3mm, inner sep=0pt, minimum size=18pt]  (par3)[below= 0.2cm of par2] {\scriptsize $\frac{\partial}{\partial t}$};

    \node[rectangle, fill = neworange, draw=black, line width=0.3mm, inner sep=3pt, minimum height = 0.7cm, minimum width = 2.5cm, rounded corners]  (eq1)[right= 0.3cm of par2] {\scriptsize $r \hat \varphi -\frac{\partial \hat \varphi}{\partial t}-\mathcal{G}_t \hat \varphi$ \: / \: $\frac{\partial F}{\partial t}-\mathcal{G}_t F$};

    \draw[edge,line width=0.3mm] (par1) -> (eq1);
    \draw[edge,line width=0.3mm] (par2) -> (eq1);
    \draw[edge,line width=0.3mm] (par3) -> (eq1);

    \node[rectangle, draw=cmublue, fit=(par1) (par2) (par3) (eq1), rounded corners, dashed, line width=0.3mm, inner sep=1mm] (block2) {};
    \node[above right] at (block2.north west){\scriptsize Physics model loss $\mathcal{L}_p$};

    \node[rectangle, fill = neworange, draw=black, line width=0.3mm, inner sep=3pt, minimum height = 0.7cm, minimum width = 2.5cm, rounded corners] (eq2)[below= 1.1cm of eq1] {\scriptsize $\hat{\varphi}(\xi,t) - \bar{\varphi}(\xi,t)$ \: / \: $\hat{F}(\xi,t) - \bar{F}(\xi,t)$};
    \node[circle, inner sep=0pt, minimum size=18pt] (par4)[below= 1.3cm of par2] {};
    \node[rectangle, draw=cmublue, fit=(par4) (eq2), rounded corners, dashed, line width=0.3mm, inner sep=1mm] (block3) {};
    \node[below right] at (block3.south west){\scriptsize Data model loss $\mathcal{L}_d$};

    \draw[edge,line width=0.3mm] (phi) -> (par1);
    \draw[edge,line width=0.3mm] (phi) -> (par2);
    \draw[edge,line width=0.3mm] (phi) -> (par3);
    \draw [arr,line width=0.3mm] (phi) |- (eq2);

    \node[rectangle, fill = newyellow, draw=black, line width=0.3mm, inner sep=0pt, minimum height = 0.7cm, minimum width = 2.5cm, rounded corners]  (loss)[below right= 0.2cm and -0.3cm of eq1] {\scriptsize  $\mathcal{L} = \omega_p \mathcal{L}_p + \omega_d \mathcal{L}_d$};
    \draw [arr,line width=0.3mm] (eq1) -| (loss);
    \draw [arr,line width=0.3mm] (eq2) -| (loss);

    \node[rectangle, fill = newgreen, draw=black, line width=0.3mm, inner sep=0pt, minimum height = 0.5cm, minimum width = 0.6cm, rounded corners]  (theta)[right= 0.3cm of loss] {\scriptsize $\boldsymbol\theta^*$};
    \path[-stealth,draw,line width=0.3mm](loss) edge node[above=0.4cm] {\scriptsize Minimize} (theta);

\end{tikzpicture}

 

%% file: Figures/AE_diagram.tex
\begin{tikzpicture}[
    node distance = 4cm,
    node/.style={circle, fill=#1, inner sep=2pt, minimum size=0.7cm}, 
    edge/.style={-stealth, #1}
]

\node[rectangle, draw, line width=0.3mm, fill = newblue, minimum height=2cm,minimum width=0.6cm,inner sep=0pt] at (0,0) (x){\scriptsize $x$};

\node[rectangle, minimum height=0.6cm,minimum width=0.6cm,inner sep=0pt] (r1)[above = 0cm of x]{\scriptsize $\mathbb{R}^n$};

\node[trapezium,draw,line width=0.3mm,trapezium angle=75, minimum width = 1.23cm, fill = newgray, 
    minimum height = 1.3cm, rotate = -90, inner sep = -3pt](encoder)at (1.2,0) {\scriptsize \rotatebox{90}{\begin{tabular}{cc} Encoder \\ $p_\sigma (x)$ \end{tabular}}};

\node[rectangle, draw, line width=0.3mm, fill = newblue, minimum height=0.6cm,minimum width=0.6cm,inner sep=0pt] at (2.5,0) (xi){\scriptsize $\xi$};

\node[rectangle, minimum height=0.6cm,minimum width=0.6cm,inner sep=0pt] (r2)[above = 0cm of xi]{\scriptsize $\mathbb{R}^k$};

\node[rectangle, minimum height=0.6cm,minimum width=0.6cm,inner sep=0pt] (l1)[below = 1cm of xi]{\scriptsize $\mathcal{L}_{C.T.}$};

\node[trapezium,draw,line width=0.3mm,trapezium angle=75, minimum width = 1.23cm, fill = newgray, 
    minimum height = 1.3cm, rotate = 90, inner sep = -3pt](decoder)at (3.8,0) {\scriptsize \rotatebox{-90}{\begin{tabular}{cc} Decoder \\ $g_\sigma (\xi)$ \end{tabular}}};

\node[rectangle, draw, line width=0.3mm, fill = newblue, minimum height=0.6cm, minimum width=0.6cm,inner sep=0pt] at (5,0) (f){\scriptsize $\hat{r}(\xi)$};

\node[rectangle, minimum height=0.6cm,minimum width=0.6cm,inner sep=0pt] (r3)[above = 0cm of f]{\scriptsize $\mathbb{R}^1$};

\node[rectangle, minimum height=0.6cm,minimum width=0.6cm,inner sep=0pt] (l2)[below = 1cm of f]{\scriptsize $\mathcal{L}_{RC}$};

\draw[edge,line width=0.3mm] (x) -> (encoder);
\draw[edge,line width=0.3mm] (encoder) -> (xi);
\draw[edge,line width=0.3mm] (xi) -> (decoder);
\draw[edge,line width=0.3mm] (decoder) -> (f);
\draw[edge,line width=0.3mm] (xi) -> (l1);
\draw[edge,line width=0.3mm] (f) -> (l2);
\end{tikzpicture}

%% file: Figures/path_integral_1kd.tex
\begin{tikzpicture}
 
\begin{axis}[
view/h=70,
view/v=50,
tick pos=left,
xlabel={\(\displaystyle \xi_2\)},
xtick = {1.2,1.6,2},
xtick style={color=black},
ylabel={\(\displaystyle \xi_1\)},
ytick = {1.2,1.6,2},
ytick style={color=black},
ztick = {0.3, 0.35, 0.4, 0.45},
zmin = 0.3, zmax = 0.48,
width=4cm,height=4cm,
font = \tiny, 
grid
]
 
\addplot3[
  mark=*,
  mark size=1pt,
  only marks,
  scatter,
  scatter/@post marker code/.code={%
  \endscope
},
] table 
{
x    y         z
1.1	1.1	0.40565765
1.1	1.2	0.40906673
1.1	1.3	0.38070724
1.1	1.4	0.41947888
1.1	1.5	0.39722215
1.1	1.6	0.327815  
1.1	1.7	0.42057389
1.1	1.8	0.4125304 
1.1	1.9	0.41043739
1.1	2.0	0.39177632
1.2	1.1	0.39057698
1.2	1.2	0.42240288
1.2	1.3	0.41134408
1.2	1.4	0.40181783
1.2	1.5	0.39592508
1.2	1.6	0.4028499 
1.2	1.7	0.38691981
1.2	1.8	0.32852271
1.2	1.9	0.37189909
1.2	2.0	0.40569639
1.3	1.1	0.4021314 
1.3	1.2	0.37647653
1.3	1.3	0.40863601
1.3	1.4	0.42108011
1.3	1.5	0.42933735
1.3	1.6	0.39541396
1.3	1.7	0.42539899
1.3	1.8	0.42294425
1.3	1.9	0.40247396
1.3	2.0	0.42562671
1.4	1.1	0.41106665
1.4	1.2	0.41515223
1.4	1.3	0.40118414
1.4	1.4	0.37091946
1.4	1.5	0.36558178
1.4	1.6	0.33109136
1.4	1.7	0.47613098
1.4	1.8	0.42098295
1.4	1.9	0.31857162
1.4	2.0	0.38683164
1.5	1.1	0.38704072
1.5	1.2	0.40674994
1.5	1.3	0.34074324
1.5	1.4	0.39740023
1.5	1.5	0.37438479
1.5	1.6	0.46459003
1.5	1.7	0.47303517
1.5	1.8	0.43332828
1.5	1.9	0.42910044
1.5	2.0	0.38717436
1.6	1.1	0.39256698
1.6	1.2	0.37941244
1.6	1.3	0.46470152
1.6	1.4	0.39593074
1.6	1.5	0.37578946
1.6	1.6	0.37891176
1.6	1.7	0.4175584 
1.6	1.8	0.3902694 
1.6	1.9	0.41664273
1.6	2.0	0.37877905
1.7	1.1	0.44378004
1.7	1.2	0.40528874
1.7	1.3	0.3980817 
1.7	1.4	0.44333453
1.7	1.5	0.40275016
1.7	1.6	0.40793259
1.7	1.7	0.42380686
1.7	1.8	0.4063768 
1.7	1.9	0.36389793
1.7	2.0	0.39502977
1.8	1.1	0.36884713
1.8	1.2	0.38433559
1.8	1.3	0.39053864
1.8	1.4	0.40482896
1.8	1.5	0.38869122
1.8	1.6	0.41995875
1.8	1.7	0.4666564 
1.8	1.8	0.41770006
1.8	1.9	0.34124801
1.8	2.0	0.43008697
1.9	1.1	0.41154597
1.9	1.2	0.37233469
1.9	1.3	0.41076619
1.9	1.4	0.41227455
1.9	1.5	0.37438566
1.9	1.6	0.3938453 
1.9	1.7	0.42059355
1.9	1.8	0.37538536
1.9	1.9	0.39709385
1.9	2.0	0.41333092
2.0	1.1	0.36769539
2.0	1.2	0.37862954
2.0	1.3	0.44736103
2.0	1.4	0.37825467
2.0	1.5	0.4228928 
2.0	1.6	0.38815098
2.0	1.7	0.35185788
2.0	1.8	0.44647427
2.0	1.9	0.38486582
2.0	2.0	0.3575396 
};
\end{axis}
 
\end{tikzpicture}

%% file: Figures/prediction_1kd.tex
\begin{tikzpicture}
 
\begin{axis}[
view/h=70,
view/v=50,
tick pos=left,
xlabel={\(\displaystyle \xi_2\)},
xtick = {1.2,1.6,2},
xtick style={color=black},
ylabel={\(\displaystyle \xi_1\)},
ytick = {1.2,1.6,2},
ytick style={color=black},
z tick label style={/pgf/number format/precision=3},
zmin=0.384, zmax=0.39,
ztick style={color=black},
width=4cm,height=4cm,
font = \tiny, 
grid
]
 
\addplot3[
  mark=*,
  mark size=1pt,
  only marks,
  scatter,
  scatter/@post marker code/.code={%
  \endscope
},
] table 
{
x    y         z
1.1	1.1	0.38822708
1.1	1.2	0.3878177 
1.1	1.3	0.3873986 
1.1	1.4	0.38696966
1.1	1.5	0.38653117
1.1	1.6	0.38608363
1.1	1.7	0.38562652
1.1	1.8	0.38516018
1.1	1.9	0.38468498
1.1	2.0	0.384201  
1.2	1.1	0.38845968
1.2	1.2	0.38805744
1.2	1.3	0.38764504
1.2	1.4	0.3872227 
1.2	1.5	0.38679066
1.2	1.6	0.38634878
1.2	1.7	0.38589743
1.2	1.8	0.3854364 
1.2	1.9	0.3849663 
1.2	2.0	0.38448724
1.3	1.1	0.3886749 
1.3	1.2	0.38827953
1.3	1.3	0.3878741 
1.3	1.4	0.3874583 
1.3	1.5	0.38703215
1.3	1.6	0.38659596
1.3	1.7	0.38615018
1.3	1.8	0.38569444
1.3	1.9	0.38522935
1.3	2.0	0.3847548 
1.4	1.1	0.3888729 
1.4	1.2	0.3884847 
1.4	1.3	0.38808578
1.4	1.4	0.38767618
1.4	1.5	0.38725623
1.4	1.6	0.3868259 
1.4	1.7	0.38638523
1.4	1.8	0.3859346 
1.4	1.9	0.38547435
1.4	2.0	0.38500416
1.5	1.1	0.38905445
1.5	1.2	0.38867337
1.5	1.3	0.3882807 
1.5	1.4	0.38787726
1.5	1.5	0.3874631 
1.5	1.6	0.3870382 
1.5	1.7	0.38660288
1.5	1.8	0.3861572 
1.5	1.9	0.38570136
1.5	2.0	0.3852356 
1.6	1.1	0.3892199 
1.6	1.2	0.38884515
1.6	1.3	0.3884591 
1.6	1.4	0.38806176
1.6	1.5	0.38765323
1.6	1.6	0.38723376
1.6	1.7	0.38680342
1.6	1.8	0.38636228
1.6	1.9	0.38591096
1.6	2.0	0.3854493 
1.7	1.1	0.38936958
1.7	1.2	0.38900158
1.7	1.3	0.38862154
1.7	1.4	0.38823003
1.7	1.5	0.387827  
1.7	1.6	0.38741255
1.7	1.7	0.38698718
1.7	1.8	0.3865506 
1.7	1.9	0.38610354
1.7	2.0	0.38564584
1.8	1.1	0.38950405
1.8	1.2	0.3891423 
1.8	1.3	0.38876835
1.8	1.4	0.3883825 
1.8	1.5	0.3879847 
1.8	1.6	0.38757548
1.8	1.7	0.3871547 
1.8	1.8	0.38672248
1.8	1.9	0.38627943
1.8	2.0	0.3858254 
1.9	1.1	0.38962328
1.9	1.2	0.38926792
1.9	1.3	0.3888999 
1.9	1.4	0.38851953
1.9	1.5	0.38812703
1.9	1.6	0.38772246
1.9	1.7	0.38730612
1.9	1.8	0.38687834
1.9	1.9	0.38643894
1.9	2.0	0.3859885 
2.0	1.1	0.38972837
2.0	1.2	0.38937902
2.0	1.3	0.38901663
2.0	1.4	0.3886415 
2.0	1.5	0.38825408
2.0	1.6	0.38785404
2.0	1.7	0.38744193
2.0	1.8	0.38701817
2.0	1.9	0.3865824 
2.0	2.0	0.38613522
};
\end{axis}
 
\end{tikzpicture}

%% file: Figures/sample_complexity.tex
\begin{tikzpicture}
\begin{axis}[
    xlabel=\footnotesize{sample complexity},
    ylabel=\footnotesize{percentage error (\%)},
    xmode = log,
    xmin=10, xmax=10^6,
    ymin=0, ymax=80,
    ytick={0,20,40,60,80},
    width=8cm,height=4cm,
    font = \scriptsize, 
            ]
\addplot[mark=x,blue] plot coordinates {
    (10^2,100*0.7675)
    (10^3,100*0.2399)
    (10^4,100*0.0680)
    (10^5,100*0.0267)
    (10^6,100*0.0132)    
};
\addlegendentry{MC w/o dimension reduction}

\addplot[color=black,mark=x,black]
    plot coordinates {
        (10^1,100*0.7392)
        (10^2,100*0.2678)
        (10^3,100*0.0797)
        (10^4,100*0.0267)
        (10^5,100*0.0117)
    };
\addlegendentry{MC w/ dimension reduction}

\addplot[color=red,mark=triangle*, mark size=3pt]
    plot coordinates {
        (10, 51.6)
        (100, 8.5)
        (1000,5)
    };
\addlegendentry{PINN w/ dimension reduction}
\end{axis}
\end{tikzpicture}

%% file: Figures/value_func_visual_a.tex
\begin{tikzpicture}
 
\begin{axis}[
tick pos=left,
xlabel={\(\displaystyle \xi_2\)},
xtick = {1.2,1.6,2},
xtick style={color=black},
ylabel={\(\displaystyle \xi_1\)},
ytick = {1.2,1.6,2},
ytick style={color=black},
ztick = {0.1,0.2,0.3,0.4},
width=4.5cm,height=4.5cm,
font = \scriptsize, 
grid
]
 
\addplot3[
  mark=*,
  mark size=1pt,
  only marks,
  scatter,
  scatter/@post marker code/.code={%
  \endscope
},
] table 
{
x    y         z
 1.1  1.1  0.399468
 1.1  1.2  0.400738
 1.1  1.3   0.37457
 1.1  1.4  0.360488
 1.1  1.5  0.328584
 1.1  1.6  0.324079
 1.1  1.7  0.298825
 1.1  1.8   0.27083
 1.1  1.9    0.2521
 1.1  2.0  0.227553
 1.2  1.1  0.389525
 1.2  1.2  0.370092
 1.2  1.3  0.352291
 1.2  1.4  0.330356
 1.2  1.5  0.309561
 1.2  1.6  0.300812
 1.2  1.7  0.276975
 1.2  1.8  0.252998
 1.2  1.9  0.245923
 1.2  2.0  0.216798
 1.3  1.1  0.351917
 1.3  1.2   0.34194
 1.3  1.3  0.325988
 1.3  1.4  0.310379
 1.3  1.5  0.286387
 1.3  1.6  0.280482
 1.3  1.7   0.25777
 1.3  1.8  0.236114
 1.3  1.9  0.224505
 1.3  2.0  0.207655
 1.4  1.1   0.34314
 1.4  1.2  0.307621
 1.4  1.3  0.311731
 1.4  1.4   0.28943
 1.4  1.5  0.275154
 1.4  1.6  0.254513
 1.4  1.7  0.243731
 1.4  1.8  0.228482
 1.4  1.9  0.210313
 1.4  2.0  0.198334
 1.5  1.1  0.308505
 1.5  1.2   0.29355
 1.5  1.3   0.28427
 1.5  1.4    0.2854
 1.5  1.5  0.258761
 1.5  1.6  0.240411
 1.5  1.7  0.210336
 1.5  1.8  0.204956
 1.5  1.9  0.189817
 1.5  2.0   0.16893
 1.6  1.1  0.293874
 1.6  1.2  0.272323
 1.6  1.3  0.257784
 1.6  1.4  0.250918
 1.6  1.5  0.238079
 1.6  1.6  0.224349
 1.6  1.7   0.19364
 1.6  1.8  0.191998
 1.6  1.9  0.169856
 1.6  2.0  0.164362
 1.7  1.1  0.256447
 1.7  1.2  0.247281
 1.7  1.3  0.236253
 1.7  1.4  0.220246
 1.7  1.5  0.216236
 1.7  1.6   0.19967
 1.7  1.7  0.188838
 1.7  1.8  0.174403
 1.7  1.9  0.163284
 1.7  2.0  0.152871
 1.8  1.1  0.249276
 1.8  1.2  0.218992
 1.8  1.3   0.21435
 1.8  1.4  0.202368
 1.8  1.5  0.182429
 1.8  1.6  0.182966
 1.8  1.7  0.172841
 1.8  1.8  0.165818
 1.8  1.9  0.146673
 1.8  2.0   0.13143
 1.9  1.1  0.214343
 1.9  1.2   0.20608
 1.9  1.3  0.194687
 1.9  1.4  0.182887
 1.9  1.5  0.172344
 1.9  1.6  0.165432
 1.9  1.7  0.150118
 1.9  1.8  0.128952
 1.9  1.9  0.128543
 1.9  2.0  0.123256
 2.0  1.1  0.194454
 2.0  1.2  0.171992
 2.0  1.3  0.165933
 2.0  1.4  0.167728
 2.0  1.5  0.152967
 2.0  1.6  0.146089
 2.0  1.7  0.148348
 2.0  1.8  0.128912
 2.0  1.9  0.118328
 2.0  2.0  0.111661
};
\end{axis}
 
\end{tikzpicture}

%% file: Figures/value_func_visual_b.tex
\begin{tikzpicture}
 
\begin{axis}[
tick pos=left,
grid,
xlabel={\(\displaystyle \xi_2\)},
xtick style={color=black},
xtick = {1.2,1.6,2},
ylabel={\(\displaystyle \xi_1\)},
ytick style={color=black},
ytick = {1.2,1.6,2},
ztick = {0,0.1,0.2,0.3,0.4,0.5,0.6},
width=4.5cm,height=4.5cm,
font = \scriptsize, 
]
 
\addplot3[
  mark=*,
  mark size=1pt,
  only marks,
  scatter,
  scatter/@post marker code/.code={%
  \endscope
},
] table 
{
x    y         z
1.1  1.1  0.566776
 1.1  1.2    0.5442
 1.1  1.3  0.521572
 1.1  1.4  0.498963
 1.1  1.5  0.476445
 1.1  1.6  0.454088
 1.1  1.7  0.431963
 1.1  1.8  0.410133
 1.1  1.9  0.388662
 1.1  2.0  0.367607
 1.2  1.1  0.542305
 1.2  1.2  0.520507
 1.2  1.3  0.498673
 1.2  1.4  0.476866
 1.2  1.5  0.455153
 1.2  1.6  0.433599
 1.2  1.7  0.412271
 1.2  1.8   0.39123
 1.2  1.9  0.370534
 1.2  2.0  0.350237
 1.3  1.1  0.517474
 1.3  1.2  0.496528
 1.3  1.3  0.475552
 1.3  1.4  0.454605
 1.3  1.5  0.433748
 1.3  1.6  0.413043
 1.3  1.7  0.392553
 1.3  1.8  0.372335
 1.3  1.9  0.352445
 1.3  2.0  0.332934
 1.4  1.1   0.49241
 1.4  1.2  0.472376
 1.4  1.3  0.452311
 1.4  1.4   0.43227
 1.4  1.5  0.412308
 1.4  1.6  0.392486
 1.4  1.7  0.372862
 1.4  1.8  0.353492
 1.4  1.9   0.33443
 1.4  2.0  0.315723
 1.5  1.1  0.467252
 1.5  1.2  0.448174
 1.5  1.3  0.429059
 1.5  1.4  0.409956
 1.5  1.5  0.390918
 1.5  1.6  0.372001
 1.5  1.7  0.353262
 1.5  1.8  0.334755
 1.5  1.9  0.316531
 1.5  2.0  0.298637
 1.6  1.1  0.442137
 1.6  1.2  0.424046
 1.6  1.3  0.405906
 1.6  1.4  0.387762
 1.6  1.5  0.369663
 1.6  1.6  0.351664
 1.6  1.7  0.333818
 1.6  1.8  0.316179
 1.6  1.9  0.298796
 1.6  2.0  0.281715
 1.7  1.1  0.417195
 1.7  1.2  0.400108
 1.7  1.3  0.382958
 1.7  1.4  0.365782
 1.7  1.5  0.348628
 1.7  1.6  0.331548
 1.7  1.7  0.314595
 1.7  1.8  0.297821
 1.7  1.9  0.281273
 1.7  2.0  0.264998
 1.8  1.1  0.392542
 1.8  1.2  0.376467
 1.8  1.3  0.360309
 1.8  1.4  0.344101
 1.8  1.5   0.32789
 1.8  1.6  0.311724
 1.8  1.7  0.295655
 1.8  1.8  0.279734
 1.8  1.9  0.264009
 1.8  2.0  0.248526
 1.9  1.1  0.368282
 1.9  1.2  0.353214
 1.9  1.3  0.338041
 1.9  1.4  0.322794
 1.9  1.5  0.307514
 1.9  1.6   0.29225
 1.9  1.7   0.27705
 1.9  1.8  0.261966
 1.9  1.9  0.247046
 1.9  2.0  0.232335
 2.0  1.1  0.344499
 2.0  1.2  0.330425
 2.0  1.3  0.316224
 2.0  1.4  0.301921
 2.0  1.5  0.287557
 2.0  1.6  0.273176
 2.0  1.7  0.258826
 2.0  1.8  0.244558
 2.0  1.9  0.230421
 2.0  2.0   0.21646
};
\end{axis}
 
\end{tikzpicture}

%% file: Figures/safe_prob_visual_a.tex
\begin{tikzpicture}
 
\begin{axis}[
tick pos=left,
xlabel={\(\displaystyle \xi_2\)},
xtick = {1.2,1.6,2},
xtick style={color=black},
ylabel={\(\displaystyle \xi_1\)},
ytick = {1.2,1.6,2},
ytick style={color=black},
ztick = {0.1,0.2,0.3,0.4,0.5},
width=4.5cm,height=4.5cm,
font = \scriptsize, 
grid
]
 
\addplot3[
  mark=*,
  mark size=1pt,
  only marks,
  scatter,
  scatter/@post marker code/.code={%
  \endscope
},
] table 
{
x    y         z
 1.1  1.1  0.522
 1.1  1.2  0.487
 1.1  1.3  0.479
 1.1  1.4  0.402
 1.1  1.5  0.376
 1.1  1.6  0.363
 1.1  1.7  0.322
 1.1  1.8   0.25
 1.1  1.9  0.232
 1.1  2.0  0.228
 1.2  1.1  0.477
 1.2  1.2  0.445
 1.2  1.3  0.427
 1.2  1.4  0.383
 1.2  1.5  0.334
 1.2  1.6  0.336
 1.2  1.7  0.293
 1.2  1.8  0.262
 1.2  1.9  0.234
 1.2  2.0  0.222
 1.3  1.1  0.462
 1.3  1.2  0.416
 1.3  1.3  0.383
 1.3  1.4  0.379
 1.3  1.5  0.326
 1.3  1.6  0.293
 1.3  1.7  0.258
 1.3  1.8  0.222
 1.3  1.9  0.217
 1.3  2.0  0.188
 1.4  1.1  0.399
 1.4  1.2  0.353
 1.4  1.3   0.37
 1.4  1.4  0.312
 1.4  1.5  0.289
 1.4  1.6  0.267
 1.4  1.7  0.226
 1.4  1.8  0.207
 1.4  1.9  0.182
 1.4  2.0  0.159
 1.5  1.1  0.355
 1.5  1.2  0.342
 1.5  1.3  0.295
 1.5  1.4  0.243
 1.5  1.5  0.249
 1.5  1.6  0.243
 1.5  1.7  0.207
 1.5  1.8  0.189
 1.5  1.9  0.189
 1.5  2.0  0.137
 1.6  1.1  0.313
 1.6  1.2  0.307
 1.6  1.3  0.267
 1.6  1.4  0.245
 1.6  1.5  0.245
 1.6  1.6  0.221
 1.6  1.7  0.176
 1.6  1.8  0.173
 1.6  1.9  0.148
 1.6  2.0  0.126
 1.7  1.1  0.275
 1.7  1.2  0.252
 1.7  1.3  0.234
 1.7  1.4  0.216
 1.7  1.5  0.198
 1.7  1.6  0.215
 1.7  1.7  0.164
 1.7  1.8  0.168
 1.7  1.9   0.13
 1.7  2.0   0.11
 1.8  1.1  0.251
 1.8  1.2  0.219
 1.8  1.3  0.177
 1.8  1.4  0.193
 1.8  1.5   0.17
 1.8  1.6  0.183
 1.8  1.7  0.142
 1.8  1.8  0.129
 1.8  1.9  0.125
 1.8  2.0  0.114
 1.9  1.1  0.197
 1.9  1.2  0.181
 1.9  1.3  0.165
 1.9  1.4  0.149
 1.9  1.5  0.167
 1.9  1.6  0.135
 1.9  1.7  0.119
 1.9  1.8  0.104
 1.9  1.9  0.103
 1.9  2.0  0.078
 2.0  1.1  0.148
 2.0  1.2  0.138
 2.0  1.3  0.125
 2.0  1.4  0.125
 2.0  1.5  0.103
 2.0  1.6  0.102
 2.0  1.7  0.095
 2.0  1.8  0.083
 2.0  1.9  0.082
 2.0  2.0  0.078
};
\end{axis}
 
\end{tikzpicture}

%% file: Figures/safe_prob_visual_b.tex
\begin{tikzpicture}
 
\begin{axis}[
tick pos=left,
xlabel={\(\displaystyle \xi_2\)},
xtick = {1.2,1.6,2},
xtick style={color=black},
ylabel={\(\displaystyle \xi_1\)},
ytick = {1.2,1.6,2},
ytick style={color=black},
ztick = {0.1,0.2,0.3,0.4},
width=4.5cm,height=4.5cm,
font = \scriptsize, 
grid
]
 
\addplot3[
  mark=*,
  mark size=1pt,
  only marks,
  scatter,
  scatter/@post marker code/.code={%
  \endscope
},
] table 
{
x    y         z
 1.1  1.1  0.387751
 1.1  1.2  0.381446
 1.1  1.3  0.366031
 1.1  1.4  0.344191
 1.1  1.5  0.318373
 1.1  1.6  0.290492
 1.1  1.7  0.261935
 1.1  1.8  0.233664
 1.1  1.9  0.206335
 1.1  2.0  0.180381
 1.2  1.1   0.34381
 1.2  1.2  0.339323
 1.2  1.3  0.326519
 1.2  1.4  0.307671
 1.2  1.5  0.284968
 1.2  1.6  0.260187
 1.2  1.7  0.234646
 1.2  1.8  0.209273
 1.2  1.9  0.184703
 1.2  2.0  0.161356
 1.3  1.1  0.301796
 1.3  1.2  0.298677
 1.3  1.3  0.288114
 1.3  1.4  0.271986
 1.3  1.5  0.252219
 1.3  1.6  0.230422
 1.3  1.7  0.207823
 1.3  1.8  0.185295
 1.3  1.9  0.163441
 1.3  2.0  0.142661
 1.4  1.1   0.26228
 1.4  1.2   0.26013
 1.4  1.3  0.251433
 1.4  1.4   0.23772
 1.4  1.5   0.22065
 1.4  1.6  0.201661
 1.4  1.7  0.181868
 1.4  1.8  0.162076
 1.4  1.9  0.142846
 1.4  2.0  0.124548
 1.5  1.1  0.225661
 1.5  1.2   0.22414
 1.5  1.3  0.216968
 1.5  1.4  0.205364
 1.5  1.5  0.190732
 1.5  1.6  0.174333
 1.5  1.7  0.157164
 1.5  1.8  0.139954
 1.5  1.9  0.123211
 1.5  2.0  0.107271
 1.6  1.1  0.192191
 1.6  1.2  0.191024
 1.6  1.3  0.185079
 1.6  1.4  0.175295
 1.6  1.5  0.162842
 1.6  1.6  0.148804
 1.6  1.7  0.134054
 1.6  1.8  0.119242
 1.6  1.9  0.104817
 1.6  2.0  0.091082
 1.7  1.1  0.162013
 1.7  1.2  0.160983
 1.7  1.3  0.156006
 1.7  1.4  0.147782
 1.7  1.5   0.13726
 1.7  1.6  0.125354
 1.7  1.7  0.112813
 1.7  1.8  0.100201
 1.7  1.9  0.087909
 1.7  2.0    0.0762
 1.8  1.1  0.135169
 1.8  1.2  0.134114
 1.8  1.3  0.129887
 1.8  1.4  0.122988
 1.8  1.5  0.114167
 1.8  1.6   0.10417
 1.8  1.7  0.093628
 1.8  1.8  0.083014
 1.8  1.9  0.072664
 1.8  2.0  0.062802
 1.9  1.1   0.11162
 1.9  1.2  0.110432
 1.9  1.3  0.106772
 1.9  1.4  0.100984
 1.9  1.5  0.093643
 1.9  1.6  0.085343
 1.9  1.7  0.076592
 1.9  1.8   0.06778
 1.9  1.9  0.059181
 1.9  2.0  0.050983
 2.0  1.1  0.091252
 2.0  1.2  0.089869
 2.0  1.3   0.08663
 2.0  1.4  0.081763
 2.0  1.5  0.075693
 2.0  1.6  0.068877
 2.0  1.7  0.061711
 2.0  1.8  0.054502
 2.0  1.9  0.047469
 2.0  2.0  0.040757
};
\end{axis}
 
\end{tikzpicture}

%% file: Figures/prediction_time.tex
\begin{tikzpicture}
\begin{axis}[
    xlabel=\footnotesize{$t$},
    ylabel=\footnotesize{absolute error},
    xmin = 0.4, xmax = 1.6,
    xtick = {0.4,0.6,0.8,1.0,1.2,1.4,1.6},
    ymin = 0, ymax =0.08,
    ytick = {0.02,0.04,0.06,0.08},
    width=8cm,height=4cm,
    font = \scriptsize,
    scaled y ticks = false,
    yticklabel style={
            /pgf/number format/fixed,
        },
]
\addplot[mark=*,black,mark size=1.5pt] plot coordinates {
    (1.5,0.0641)
    (1.4,0.02969)
    (1.3,0.0136)
    (1.2,0.0097)
    (1.1,0.0132)    
    (1,0.02005)   
    (0.9,0.02835)   
    (0.8,0.03566)
    (0.7,0.04163)   
    (0.6,0.04558)   
    (0.5,0.04903)           
};
\end{axis}
\end{tikzpicture}

%% file: Figures/pinn_epoch.tex
\begin{tikzpicture}
\begin{axis}[
    axis y line*=left,
    xlabel=\footnotesize{\# epoch},
    ylabel=\footnotesize{percentage error},
    xmin = 10000, xmax = 60000,
    xtick = {10000,20000,30000,40000,50000,60000},
    ymin = 0, ymax =12,
    ytick = {0,2,4,6,8,10,12},
    yticklabel={$\pgfmathprintnumber{\tick}\%$},
    width=8cm,height=4cm,
    scaled y ticks = false,
    yticklabel style={
            /pgf/number format/fixed,
        },
    font = \scriptsize
]
\addplot [mark=*, black, mark size=1.5pt]table[x=x,y=y] {pinn_epoch.dat};\label{plot_one}
\end{axis}


\begin{axis}[
    axis y line*=right,
    ylabel=\footnotesize{absolute error},
    xmin = 10000, xmax = 60000,
    xtick = {10000,20000,30000,40000,50000,60000},
    ymin = 0.01, ymax =0.04,
    ytick = {0,0.01,0.02,0.03,0.04},
    width=8cm,height=4cm,
    scaled y ticks = false,
    yticklabel style={
            /pgf/number format/fixed,
        },
    font = \scriptsize
]
\addplot [mark=*, blue, mark size=1.5pt]table[x=x,y=err] {pinn_epoch.dat};\label{plot_two}
\addlegendimage{/pgfplots/refstyle=plot_one}\addlegendentry{percentage error}
\addlegendimage{/pgfplots/refstyle=plot_two}\addlegendentry{absolute error}
\end{axis}
\end{tikzpicture}

%% file: Tables/pinn_epoch_table.tex

\begin{table*}[t]
\small
\centering
\begin{tabular}{c|cccccc}
\Xhline{3\arrayrulewidth}
\# Epoch($\times 10^{4}$)          & 1                          & 2                          & 3      & 4   & 5   &6                   \\ \hline
Prediction Error ($\%$) & 7.104& 5.322  & 5.475  & 5.940 & 4.733 & 4.809 \\\hline
Absolute Error ($\times 10^{-2}$) & 2.588 & 2.009 &  2.065 & 2.252 &1.815&1.837\\\Xhline{3\arrayrulewidth}
\end{tabular}
\caption{PINN with different epochs for training}
\label{tab:pinn_epoch}
\end{table*}

%% file: Figures/pinn_pde.tex



\begin{tikzpicture}
\begin{axis}[
    axis y line*=left,
    xlabel=\footnotesize{\# domain},
    ylabel=\footnotesize{percentage error},
    xmin = 0, xmax = 2500,
    xtick = {0,200,400,600,1200,2400},
    ymin = 0, ymax =12,
    ytick = {0,2,4,6,8,10,12},
    yticklabel={$\pgfmathprintnumber{\tick}\%$},
    width=8cm,height=4cm,
    scaled y ticks = false,
    yticklabel style={
            /pgf/number format/fixed,
        },
    font = \scriptsize
]
\addplot [mark=*, black, mark size=1.5pt]table[x=x,y=y] {pinn_pde.dat};\label{plot1}
\end{axis}


\begin{axis}[
    axis y line*=right,
    ylabel=\footnotesize{absolute error},
    xmin = 0, xmax = 2500,
    xtick = {0,200,400,600,1200,2400},
    ymin = 0.01, ymax =0.04,
    ytick = {0,0.01,0.02,0.03,0.04},
    width=8cm,height=4cm,
    scaled y ticks = false,
    yticklabel style={
            /pgf/number format/fixed,
        },
    font = \scriptsize
]
\addplot [mark=*, blue, mark size=1.5pt]table[x=x,y=err] {pinn_pde.dat};\label{plot2}
\addlegendimage{/pgfplots/refstyle=plot1}\addlegendentry{percentage error}
\addlegendimage{/pgfplots/refstyle=plot2}\addlegendentry{absolute error}
\end{axis}
\end{tikzpicture}

%% file: Tables/pinn_pde_table.tex

\begin{table*}[t]
\small
\centering
\begin{tabular}{ccccccccccccc}
\Xhline{3\arrayrulewidth}
\multicolumn{1}{l|}{\# Domain}           & 100                          & 200                          & 300                         & 400                & 500                            & 600     &1200       &2400                               \\ \hline
\multicolumn{1}{l|}{Prediction Error ($\%$)} & 5.498 & 5.391 & 5.749  & 5.927 & 5.478 & 5.283 & 5.153 & 4.992\\ \hline   
\multicolumn{1}{l|}{Absolute Error ($\times 10^{-2}$)} & 2.082 & 2.023 & 2.171 & 2.221 &2.058 &1.989 &1.962 &1.894 \\ \Xhline{3\arrayrulewidth}
\end{tabular}
\caption{PINN with different numbers of PDE constraint grid points}
\label{tab:pinn_pde}
\end{table*}

%% file: Tables/ae_table1.tex
\begin{table*}[!ht]
\small
\centering
\begin{tabular}{ccccccccccccc}
\Xhline{3\arrayrulewidth}
\multicolumn{1}{l|}{Network Architecture}           & [16]  & [64] & [128]                           \\ \hline
\multicolumn{1}{l|}{Feature Representation Error ($\%$)} & 0.248 & 0.250 & 0.251 \\ \hline   
\multicolumn{1}{l|}{Cost Reconstruction Error ($\%$)} & 0.901 & 0.984 & 1.003 \\ \Xhline{3\arrayrulewidth}

\end{tabular}
 \caption{Autoencoder-like network architecture for $n_{\text{layer}}=1$}
\label{tab:n_layer = 1}
\end{table*}

%% file: Tables/ae_table2.tex
\begin{table*}[!ht]
\small
\centering
\begin{tabular}{ccccccccccccc}
\Xhline{3\arrayrulewidth}
\multicolumn{1}{l|}{Network Architecture}           & [16, 64, 16]  & [64, 64, 64] & [64, 128, 64]                           \\ \hline
\multicolumn{1}{l|}{Representation Error ($\%$)} & 0.112 & 0.394 & 0.119 \\ \hline   
\multicolumn{1}{l|}{Cost Reconstruction Error ($\%$)} & 0.868 & 0.880 & 0.371 \\ \Xhline{3\arrayrulewidth}

\end{tabular}
\caption{Autoencoder-like network architecture for $n_{\text{layer}}=3$}
\label{tab:n_layer = 3}
\end{table*}

%% file: Tables/ae_table3.tex
\begin{table*}[!ht]
\small
\centering
\begin{tabular}{ccccccccccccc}
\Xhline{3\arrayrulewidth}
\multicolumn{1}{l|}{Network Architecture}           & [16, 64, 128, 64, 16]  & [64, 64, 128, 64, 64] & [64, 128, 128, 128, 64]                     \\ \hline
\multicolumn{1}{l|}{Representation Error ($\%$)} & 0.007 & 0.011 & 0.009 \\ \hline   
\multicolumn{1}{l|}{Cost Reconstruction Error ($\%$)} & 0.120 & 0.099 & 0.129 \\ \Xhline{3\arrayrulewidth}

\end{tabular}
\caption{Autoencoder-like network architecture for $n_{\text{layer}}=5$}
\label{tab:n_layer = 5}
\end{table*}